\documentclass[10pt]{article}
\usepackage{amsmath}
\usepackage{amssymb}
\usepackage{graphicx}
\usepackage{subfigure}
\usepackage{epstopdf}

\usepackage{setspace}

\def\Rv{{\mathbf{R}}}
\def\la{\langle}
\def\ra{\rangle}
\def\d{\delta}
\def\rv{{\mathbf{r}}}
\def\t{\tau}
\def\blp{\bigg(}
\def\brp{\bigg)}
\def\N{{{\cal{N}}}}

\def\k{\kappa}
\def\e{\epsilon}
\def\a{\alpha}

\title{Kinetics of loop formation in polymer chains}
\author{Ngo Minh Toan$^{(a,b)}$, Greg Morrison$^{(a,c)}$ Changbong Hyeon$^{d}$, and D. Thirumalai$^{(a,e)}$}

\begin{document}


\begin{centering}

{\LARGE{\bf{Kinetics of Loop Formation in Polymer Chains}}}\\
\vspace{.3in}

{\large{Ngo Minh Toan$^{(a,b)}$, Greg Morrison$^{(a,c)}$ Changbong Hyeon$^{(d)}$, and D. Thirumalai$^{(a,e)}$\\}}
\vspace{.3in}

(a):  Biophysics Program, Institute for Physical Science and Technology, University of Maryland at College Park, College Park, Maryland, 20742\\
(b):  Institute of Physics and Electronics, 10-Dao Tan, Hanoi, Vietnam (on leave)\\
(c):  Department of Physics, University of Maryland at College Park, College Park, Maryland, 20742\\
(d):  Center for Theoretical Biological Physics, University of California at San Diego, La Jolla, California, 92093\\
(e):  Department of Chemistry and Biochemistry, University of Maryland at College Park, College Park, Maryland, 20742

\end{centering}

\vspace{.2in}

\begin{abstract}
We investigate the kinetics of loop formation in flexible ideal polymer chains (Rouse model), and polymers in good and poor solvents.  We show for the Rouse model, using a modification of the theory of Szabo, Schulten, and Schulten, that the time scale for cyclization is $\t_c\sim \t_0 N^2$ (where $\t_0$ is a microscopic time scale and $N$ is the number of monomers), provided the coupling between the relaxation dynamics of the end-to-end vector and the looping dynamics is taken into account. The resulting analytic expression fits the simulation results accurately when $a$, the capture radius for contact formation, exceeds $b$, the average distance between two connected beads.  Simulations also show that, when $a < b$, $\t_c\sim N^{\a_\t}$, where $1.5<{\a_\t}\le 2$ in the range $7<N<200$ used in the simulations.  By using a diffusion coefficient that is dependent on the length scales $a$ and $b$ (with $a<b$), which captures the two-stage mechanism by which looping occurs when $a < b$, we obtain an analytic expression for $\tau_c$ that fits the simulation results well.  The kinetics of contact formation between the ends of the chain are profoundly affected when interactions between monomers are taken into account.  Remarkably, for $N < 100$ the values of $\tau_c$ decrease by more than two orders of magnitude when the solvent quality changes from good to poor.   Fits of the
simulation data for $\t_c$ to a power law in $N$ ($\t_c\sim N^{\a_\t}$) show that $\a_\t$ varies from about 2.4 in a good solvent to about 1.0 in poor solvents.  The effective exponent $\a_\t$ decreases as the strength of the attractive monomer-monomer interactions increases. Loop formation in poor solvents, in which the polymer adopts dense, compact globular conformations, occurs by a reptation-like mechanism of the ends of the chain. The time for contact formation between beads that are interior to the chain in good solvents changes non-monotonically as loop length varies. In contrast, the variation is monotonic in poor solvents. The implications of our results for contact formation in polypeptide chains, RNA, and single stranded DNA are briefly outlined.
\end{abstract}

\section{Introduction}

Contact formation (cyclization) between the ends of a long polymer has been intensely studied both experimentally~\cite{WinnikBook,FloryBook} and theoretically~\cite{WF,SSS,Doi_Chem_Phys,Sokolov,Debnath,Guo,ToanPRL06}.  More recently, the kinetics of loop formation has become increasingly important largely because of its relevance to DNA looping \cite{Widom,DNA_Cyc_2} as well as protein \cite{Eaton_1,Eaton_2,IChang,Gray07JPCB,poor_expt_1,Sahoo06,Hagen01,Sauer03} and RNA  folding \cite{DTCB}. The ease of cyclization in DNA, which is a measure of its intrinsic flexibility \cite{DNA_Cyc_2,DNA_Cyc_1}, is important in gene expression and interactions with proteins and RNA.  In addition, the formation of contacts between residues  (nucleotides) near the loop \cite{Guo} may be the key nucleating event in protein (RNA) folding. For these reasons, a number of experiments have probed the dependence of the rates of cyclization in proteins \cite{Eaton_1,Eaton_2,poor_expt_2} and RNA~\cite{woodside_1,woodside_2} as a function of loop length. The experimental reports, especially on rates of loop formation in polypeptides and proteins,  have prompted a number of theoretical studies \cite{Debnath,Portman,Bovine} that build on the pioneering treatments due to Wilemski and Fixman \cite{WF} (WF) and Szabo, Schulten, and Schulten \cite{SSS} (SSS). The WF formalism determines the loop closure time $\tau_c$ by solving the diffusion equation in the presence of a sink term. The sink function accounts for the possibility that contact between the ends of a polymer chain occurs whenever they are in proximity.  The time for forming a loop is related to a suitable time integral of the sink-sink correlation function.

In an important paper, SSS developed a much simpler theory to describe the dependence of the rate of end-to-end contact formation in an ideal chain on the polymer length $N$. The SSS approximation \cite{SSS} describes the kinetics of contact formation between the ends of the chain as a diffusion process in an effective potential that is derived from the probability distribution $P(\Rv_{ee})$ of finding the chain ends with the end-to-end distance $\Rv_{ee}$.  More recently, such an approach has been adopted to obtain rates of folding of proteins from a free energy surface expressed in terms of an appropriately chosen reaction coordinate~\cite{Socci}.  The validity of using the dynamics in a potential of mean force, $F(\Rv_{ee})\sim -k_B T\log[P(\Rv_{ee})]$, to obtain $\t_c$ hinges on local equilibrium being satisfied, i.e. that all processes except the one of interest must occur rapidly.  In the case of cyclization kinetics in simple systems (Rouse model or self-avoiding polymer chains), the local equilibrium approximation depends minimally on the cyclization time $\t_c$, and the internal chain relaxation time $\t_R$.  In the limit $\t_c/\t_R\gg 1$, one can envision the motions of the ends as occurring in the effective free energy $F(\Rv_{ee})$, because the polymer effectively explores the available volume before the ends meet.  By solving the diffusion equation for an ideal chain for which $F(\Rv_{ee})\sim 3k_B T \Rv_{ee}^2/2\bar R_{ee}^2$, with $\bar R_{ee}\sim b\sqrt{N}$, where $b$ is the monomer size, subject to absorbing boundary conditions, SSS showed that the mean first passage time for contact formation ($\sim \t_c$) is $\t_{SSS}\sim \t_0 N^{\frac{3}{2}}$, where $\t_0$ is a microscopic time constant (see below, eq. (\ref{SSSOld})).

The simplicity of the SSS result, which reduces contact-formation kinetics to merely computing $P(\Rv_{ee})$, has resulted in its widespread use to fit experimental data on polypeptide chains~\cite{Eaton_1,Eaton_2,poor_expt_2}. The dependence of $\t_c$ on $N$ using the SSS theory differs from the WF predictions.  In addition, simulations also show that $\t_c$ deviates from the SSS prediction \cite{PastorZwanzig, SakataDoi, ChenTsao, Alexi}. The slower dependence of $\t_{SSS}$ on $N$ can be traced to the failure of the assumption that all internal chain motions occur faster than the process of interest.  The interplay between $\t_c$ and $\t_R$, which determines the validity of the local equilibrium condition, can be expressed in terms of well known exponents that characterize equilibration and relaxation properties of the polymer chain.  Comparison of the conformational space explored by the chain ends and the available volume prior to cyclization~\cite{deGennes76} allows us to express the validity of the local equilibrium in terms of $\theta=(d+g)/z$, where $d$ is the spatial dimension, $g$ is the des Cloizeaux correlation hole exponent that accounts for the behavior of $P(\Rv_{ee})$ for small $\Rv_{ee}$, i.e., $P(\Rv_{ee})\sim \Rv_{ee}^g$, and $z$ is the dynamical scaling exponent ($\tau_R \sim \bar R_{ee}^z$).  Additional discussions along these lines are given in Appendix A.  The SSS assumption is only a valid provided $\theta >1$ \cite{CBDT}. For the Rouse chain in the freely draining limit ($\nu = 1/2, g=0, d=3, z= 4$) gives $\theta <1$, and hence $\tau_c$ will show deviation from the SSS predictions for all $N$.

The purpose of this paper is two fold. (i) The theory based on the WF formalism and simulations
show the closure time $\tau_{WF} \sim \left< {
\Rv}_{ee}^2\right>/D_c \sim N^{1+2\nu}$ ($\nu
\approx 3/5$ for self-avoiding walk and $\nu=1/2$ for the Rouse chain) where $D_c$ is a diffusion constant. We show that the WF
result for Rouse chains, $\tau_{WF}$, can be obtained within the SSS framework,
provided an effective diffusion constant that accounts for the relaxation dynamics of the
ends of the chains is used instead of the monomer diffusion coefficient $D_0$. Thus, the simplicity of the SSS approach can be preserved while recovering the expected scaling result \cite{WF,Doi_Chem_Phys} for the dependence of $\t_c$ on $N$.  (ii) The use of the Rouse model may be appropriate for
polymers or polypeptide chains near $\Theta$-conditions. In both good and poor solvents interactions between monomers determine the statics and dynamics of the polymer chains. The chain will swell in good solvents ($\nu\approx 3/5$) whereas in poor
solvents, polymers and polypeptide chains adopt compact globular conformations. In these
situations, interactions between the monomers or the amino acid residues affect $\tau_c$. The monomer-monomer interaction energy
scale, $\e_{LJ}$, leading to the chain adopting a swollen or globular conformation
influences both $\nu$ and the chain relaxation
dynamics, and hence affects $\tau_c$. Because analytic theory in this situation is difficult, we
provide simulation results for $\tau_c$ as a function of $\e_{LJ}$ and for $10< N \le100$.

\section{Derivation of $\t_{WF}$ for the Rouse Model using the SSS Approximation}

The Rouse chain consists of $N$ beads, with two successive beads connected by a harmonic potential that keeps them at an average separation $b$, (the Kuhn length).  Contact formation between the chain ends can occur only if fluctuations result in monomers 1 and $N$ being within a capture radius $a$.  In other words, the space explored by the chain ends must overlap within the contact volume $\sim a^3$. There are three relevant time scales that affect loop closure dynamics; namely, $\t_0\approx b^2/D_0$, the fluctuation time scale of a single monomer, $\t_{ee}$, the relaxation time associated with the fluctuations of the end-to-end distance, and $\t_R$, the relaxation time of the entire chain.  Clearly, $\t_{ee}<\t_c\sim \t_R$. Because loop formation can occur only if the ends can approach each other, processes that occur on time scale $\tau_{ee}$ must be coupled to looping dynamics.  We obtain the scaling of $\t_c$ with $N$, found using the WF approximation, from the SSS formalism using a diffusion constant evaluated on the time scale $\t_{ee}$.

\subsection{Fluctuations in $\Rv_{ee}$}

The Langevin equation for a Gaussian chain is \cite{EdwardsBook}
\begin{eqnarray}
\gamma \frac{\partial\rv(s,t)}{\partial t}=-\frac{\d H_0[\rv(s,t)]}{\d \rv(s,t)}+\vec{\eta}(s,t),\label{Langevin}
\end{eqnarray}
where $\vec{\eta}(s,t)$ a white noise force with $\la \vec{\eta}(s,t)\ra=0$, $\la \vec{\eta}(s,t)\cdot\vec{\eta}(s',t')\ra=6\gamma k_B T \d(t-t')\d(s-s')$.  $\gamma$ is the friction coefficient, and $D_0=k_B T/\gamma$ is the microscopic diffusion coefficient.  By writing $\rv(s,t)=\rv_0+2\sum_{n=1}^{N-1}\rv_n(t)\cos(n\pi s/N)$, the Gaussian Hamiltonian $H_0$ becomes
\begin{eqnarray}
H_0=\frac{3}{2b^2}\int_0^N ds \blp\frac{\partial\rv(s,t)}{\partial s}\brp^2=\frac{3}{2Nb^2}\sum_{n}{} n^2\pi^2\rv^2_n(t).
\end{eqnarray}
The equation of motion for each mode
\begin{eqnarray}
\dot\rv_n(t)=-\frac{3n^2\pi^2D_0}{N^2 b^2}\rv_n(t)+\vec{\eta}_n(t).\label{eqmodes}
\end{eqnarray}
can be solved independently. The solutions naturally reveal the time scale for global motions of the chain, $\t_R=N^2 b^2/3D_0\pi^2\sim N^2 b^2/D_0$.  We note that $\tau_R$ is much larger than the relevant time scale for internal motions of the monomers, $\t_1\approx b^2/D_0$ for large $N$.  Eq. (\ref{eqmodes}) can be solved directly, and the fluctuations in the end-to-end distance $\Rv_{ee}$ are given by
\begin{eqnarray}
\la\d\Rv_{ee}^2(t)\ra=16 Nb^2\sum_{n\mbox{ odd}}\frac{N^2}{n^4\pi^4}\sin^2\blp\frac{n\pi}{N}\brp\ \blp 1-e^{-n^2t/\t_R}\brp.\label{Ree}
\end{eqnarray}
with $\la\d\Rv_{ee}^2(t)\ra\equiv\la[\Rv_{ee}(t)-\Rv_{ee}(0)]^2\rangle$.  The details of the calculation leading to eq. (\ref{Ree}) are given in Appendix C.  If we define an effective diffusion constant using
\begin{eqnarray}
D(t)=\frac{\la\d\Rv_{ee}^2(t)\ra}{6t},\label{dt}
\end{eqnarray}
then $D(0)=2D_0$, as is expected for the short time limit \cite{SSS,ChenTsao}.  On time scales on the order of $\t_R$, we find $D(\t_R)\sim D_0/N$, which is identical to the diffusion constant for the center of mass of the chain \cite{EdwardsBook}.  This is the expected result for the diffusion constant for global chain motion.

\subsection{The Effective Diffusion Constant}

The theory of Szabo, Schulten, and Schulten \cite{SSS} (SSS) determines the loop closure time by replacing the difficult polymer problem, having many degrees of freedom, with a single particle diffusing in a potential of mean force.  With this approximation, $\t_c$, which can be related to the probability that the contact is not formed (see Appendix B for more details), becomes
\begin{eqnarray}
\t_c=\frac{1}{\N}\int_a^{Nb} dr\frac{1}{D(r)P(r)}\blp\int_r^{Nb} dr' P(r')\brp^2\ +\frac{1}{\k\  \N P(a)},\label{SSS}
\end{eqnarray}
where loop closure occurs when $|\Rv_{ee}|= a$, the closure (or capture) radius, with rate $\k$, $P(r)$ is the equilibrium end-to-end distribution of the chain, and $\N=\int_a^{Nb} dr\ P(r)$. In this paper, we will consider only a chemically irreversible process, with the binding rate constant $\k\to\infty$.  In the case of the non-interacting Gaussian chain, $P(r)\sim r^2\exp(-3r^2/2Nb^2)$.  If $D(r)\sim D_0$ is a constant, it is simple to show \cite{SSS} that, for large $N$, the loop closure time is
\begin{eqnarray}
\tau_{SSS}\approx \frac{1}{3}\sqrt{\frac{\pi}{6}}\,\frac{N^{\frac{3}{2}}b^3}{D_0a}.\label{SSSOld}
\end{eqnarray}
The scaling of $\tau_{SSS}$ with $N$ given in eq. (\ref{SSSOld}) disagrees with other theories \cite{WF,Debnath} and numerous simulations \cite{PastorZwanzig,SakataDoi, ChenTsao, Alexi} that predict $\t_c\sim N^2$ for $Nb^2\gg a^2$ and $a\ge b$.  It has been noted \cite{Portman,CBDT} that the SSS theory may be a lower bound on the loop closure time for a freely draining Gaussian chain, and that an effective diffusion coefficient that is smaller than $D_0$ is required to fit the simulated \cite{Portman} and experimental \cite{Loop_expt_denature} data using $\tau_{SSS}$.  Physically, the use of a smaller diffusion constant is needed because contact formation requires fluctuations that bring $|\Rv_{ee}|$ within the capture radius $a$, a mechanism in which $\t_{ee}$ plays a crucial role.

As noted by Doi \cite{Doi_Chem_Phys}, the relevant time scale for loop closure is not simply the global relaxation time.  The fluctuations in $\Rv_{ee}$ are given not only by the longest relaxation time, but also from important contributions that arise from higher modes.  This gives rise to the differences between the Harmonic Spring and Rouse models \cite{Doi_Chem_Phys, SakataDoi}. In the Harmonic Spring model, the chain is replaced with only one spring which connects the two ends of the chain. The spring constant is chosen to reproduce the end-to-end distribution function. The higher order modes give rise to excess fluctuations on a scale $\sim 0.4 \sqrt{N} b=R'$, and their inclusion is necessary to fully capture the physics of loop closure.  In the approximation of a particle diffusing in an effective potential (as in the SSS theory), this time scale is simple to determine.  If we consider only the $x$ component of $\Rv_{ee}$, we can treat it as a particle diffusing in a potential $U_{eff}(R_x)=3R_x^2/2Nb^2-{O}(1)$, with diffusion constant $D=2D_0$.  In this case, we find
\begin{eqnarray}
\la\d R_x^2(t)\ra=\frac{2}{3}Nb^2\blp 1-e^{-t/\t_{ee}}\brp,
\end{eqnarray}
and $\la\Rv^2_{ee}(t)\ra=3\la\d R_x^2(t)\ra$, giving the natural end-to-end relaxation time $\t_{ee}=Nb^2/6D_0$.  Because we have evaluated $\t_{ee}$ using diffusion in an effective potential, the dependence of $\t_{ee}$ on $N$ should be viewed as a mean field approximation.

We can determine the effective diffusion constant on the time scale $\t_{ee}$, which includes the relaxation of $\Rv_{ee}(t)$ at the mean field level.  We define the effective diffusion constant as
\begin{eqnarray}
D_{ee}=\lim_{t\sim \t_{ee}}\frac{\la\d\Rv_{ee}^2(t)\ra}{6t}\label{Deval}.
\end{eqnarray}
with $\la\d\Rv^2_{ee}(t)\ra$ in eq. (\ref{Ree}), which includes all of the modes of the chain, and not simply the lowest one.  Noting that $\t_{ee}/\t_R\sim N^{-1}\ll1$ for large $N$, we can convert the sum in eq. (\ref{Ree}) into an integral:
\begin{eqnarray}
\la\d\Rv_{ee}^2\ra&\approx&
\frac{2\sqrt{2}}{\pi}\,N^{\frac{3}{2}}b^2\int_0^\infty dx\ \frac{\sin^2(bx/\sqrt{3D_0t})}{x^4}\blp1-e^{-x^2}\brp\label{Rfluct}\\
&\approx&8b\sqrt{\frac{3D_0t}{\pi}}.
\end{eqnarray}
In particular, for $t\approx \t_{ee}/2=Nb^2/12D_0$,
\begin{eqnarray}
D_{ee}\approx \frac{8\,D_0}{\sqrt{N\pi}}-\frac{16D_0}{3N}+O(N^{-\frac{3}{2}}).\label{DN}
\end{eqnarray}
We expect these coefficients to be accurate to a constant on the order of unity.  The effective diffusion constant $D_{ee}$ takes the higher order modes of the chain into account, and should capture the essential physics of the loop closure.  In other words, on the time scale $\t_{ee}$, resulting in $D_{ee}\sim N^{-\frac{1}{2}}$, the monomers at the chain ends are within a volume $\sim a^3$, so that contact formation is possible.

Substituting $D_{ee}$ into eq. (\ref{SSSOld}) gives
\begin{eqnarray}
\t_c\approx \frac{N^2 b^3\pi}{24\sqrt{6}\,D_0 a} \sim\t_{WF},\label{SSS_new}
\end{eqnarray}
in the limit of large $N$.  Thus, within the SSS approximation, the $N^2$ dependence of $\t_c$ may be obtained, provided the effective diffusion constant $D_{ee}$ is used.  The importance of using a diffusion constant that takes relaxation dynamics of $\Rv_{ee}$ into account has also been stressed by Portman \cite{Portman}.  The closure time in eq. (\ref{SSS_new}) depends on the capture radius as $a^{-1}$, which disagrees with the $a$-independent prediction of Doi \cite{Doi_Chem_Phys}.  In addition, eq. (\ref{SSS_new}) does not account for the possibility of $\t_c\sim N^{\a_\t}$, with $1.5< {\a_\t} < 2$, as observed with simulations by Pastor et. al. \cite{PastorZwanzig} when the capture radius $a<b$.  Both of these discrepancies are discussed in the next section, using insights garnered from simulations.



\subsection{Simulations of Loop Closure Time for Freely Jointed Chains}

In order to measure $\Rv_{ee}(t)$ and $\t_c$ for a non-interacting Freely Jointed chain, we have performed extensive Brownian dynamics simulations.
We model the connectivity of the chain using the Hamiltonian
\begin{eqnarray}
\beta H=\frac{k_s}{2}\sum_{i=1}^{N}\blp 1-\frac{|\rv_{i+1}-\rv_i|}{b_0}\brp^2,\label{Discrete_Ham}
\end{eqnarray}
with $b_0=0.38$nm, and a spring constant $k_s=100$.  We note that $\la (\rv_{i+1}-\rv_i)^2\ra^{\frac{1}{2}}\approx 0.39$nm for this Hamiltonian, which we take as the Kuhn length $b$ when fitting the data.   For large $N$, the differences between the FJC and Rouse models are not relevant, and hence the scaling of $\t_c$ with $N$ for these two models should be identical.  The microscopic diffusion coefficient was taken as $D_0= 0.77$nm$^2$/ns. The equations of motion in the overdamped limit were integrated using the Brownian dynamics algorithm \cite{SOP}, with a time step of $\Delta t=10^{-4}$ns.  The end-to-end distribution $P(r)$ is easily computed for the model in eq. (\ref{Discrete_Ham}), giving the expression for large $k_s$
\begin{eqnarray}
P(r)=2r\int_0^\infty dq\, q\sin(qr)\blp\frac{b_0 q\cos(b_0 q)+k_s\sin(b_0 q)}{b_0 q(1+k_s)}e^{-b_0^2 q^2/k_s^2}\brp^{N-1},\label{PHarm}
\end{eqnarray}
which must be numerically integrated.

In our simulations, we computed the mean first passage time directly.  We generated the initial conditions by Monte Carlo equilibration.  Starting from each equilibrated initial configuration, the equations of motion were integrated  until $|\Rv_{ee}|\le a$ for the first time, with the first passage time computed for multiple values of $N$ and $a$.  The loop closure time $\t_c$ was identified with the mean first passage time, obtained by averaging over 400 independent trajectories.  For comparison with the analytic theory, we calculated the modified SSS first passage time, with $P(r)$ given in eq. (\ref{PHarm}), and $D_{ee}$ given in eq. (\ref{DN}).  The results are shown in Fig.~\ref{fig1}.  We find that the behavior of $\t_c$ depends strongly on the ratio $a/b$.

$a\ge b:$ For $N\lesssim 100$ and $a\ge b$, we find that the modified SSS theory using the effective diffusion constant $D_{ee}$ in eq. (\ref{DN}) gives an excellent fit to the data, as a function of both $N$ and $a$ (Fig.~\ref{fig1}(A)).  Thus, modeling the loop closure process as a one-dimensional diffusive process in a potential of mean force is appropriate, {\em so long as a diffusion coefficient that takes the dynamics of the chain ends into account is used}.

For $N\gtrsim100$ and $a\ge b$, we notice significant deviations in the data from the theoretical curves.  The data points appear to converge as $a$ is varied for large $N$, suggesting the emergence of Doi's \cite{Doi_Chem_Phys} predicted scaling of $\t_c\sim N^2 a^0$.  This departure from the predictions of eq. (\ref{SSS_new}) suggests that the one-dimensional mean field approximation, which gives rise to the $a$ dependence of $\t_c$, breaks down.  Even our modified theory, which attempts to include fluctuations in $\Rv_{ee}$ on a mean field level leading to $D_{ee}$, cannot accurately represent the polymer as a diffusive process with a single degree of freedom for large $N$.  In this regime, the many degrees of freedom of the polymer must be explicitly taken into account, making the WF theory \cite{WF} more appropriate.

$a< b:$ The condition $a<b$ is non-physical for a Freely Jointed Chain with excluded volume, and certainly not relevant for realistic flexible chains in which excluded volume interaction between monomers would prevent the approach of the chain ends to distances less than $b$. (Note that for Wormlike Chains, with the statistical segment $l_p > b$, the equivalent closure condition $a < l_p$ is physically realistic. The effect of chain stiffness, which has been treated elsewhere~\cite{CBDT}, is beyond the scope of this article.) In this case (Fig. \ref{fig1}(B)), we find $\t_c\sim N^{\a_\t}$, with $1.5<{\a_\t}< 2$, in agreement with the simulation results of Pastor et. al. \cite{PastorZwanzig}.  In deriving $D_{ee}$, we assumed, as did Doi \cite{Doi_Chem_Phys}, that the relaxation of the end-to-end vector is rate limiting.  Once $|\Rv_{ee}|\sim R'\approx 0.4 \sqrt{N}b$, we expect the faster internal motions of the chain will search the conformational space rapidly, so that $\t_c$ is dominated by the slower, global motions of the chain (i.e. it is diffusion limited).  This assumption breaks down if $a\ll b$, because the endpoints must search longer for each other using the rapid internal motions on a time scale $b^2/D_0$.  In the limit of small $a$, the memory of the relaxation of the ends of the chain is completely lost.  Our derivation of $D_{ee}$, using a mean field approach, can not accurately describe the finer details when the endpoints search for each other over very small length scales, and hence our theory must be modified in this regime.

We view the loop closure for small $a$ ($<b$) as a two step process (Fig.~\ref{fig:kappa}), with the first being a reduction in $|\Rv_{ee}|\sim b$.  The first stage is well modeled by our modified SSS theory (see Fig. \ref{fig1}(A)) using the effective diffusion coefficient in eq. (\ref{DN}).  The second stage involves a search for the two ends within a radius $b$, so that contact can occur whenever $|\Rv_{ee}|= a<b$.  The large scale relaxations of the chain are not relevant in this regime.  We therefore introduce a scale-dependent diffusion coefficient
\begin{eqnarray}
D_{ee}(x)\approx\left\{\begin{array}{cc}8D_0/\sqrt{N\pi} & x> b \\ 2D_0 & x\le b\end{array}\right.\label{DsmallA}.
\end{eqnarray}
Substitution of eq. (\ref{DsmallA}) into eq. (\ref{SSS}) with $P(r)$ given by eq. (\ref{PHarm}) yields, for $a\le b$,
\begin{eqnarray}
\t_c(a)\approx \frac{N^2 b^2\pi}{24\sqrt{6}\,D_0}+\frac{N^{3/2}b^2(b-a)\sqrt{\pi}}{6\sqrt{6}D_0a}\label{TauSmallA}.
\end{eqnarray}
In Fig. \ref{fig1}(B), we compare the predictions of eq. (\ref{TauSmallA}) for the closure time to the simulated data for $a\le b$.  The fit is excellent, showing that the simple scale-dependent diffusion coefficient (eq.~(\ref{DsmallA})), that captures the two stage mechanism of cyclization when $a<b$, accurately describes the physics of loop closure for small $a$.  By equating the two terms in eq. (\ref{TauSmallA}), we predict that the $N^{3/2}$ scaling will begin to emerge when $N\lesssim 16 b^2 (a/b-1)^2/a^2\pi$.  This upper bound on $N$ is consistent with the predictions of Chen et. al. \cite{ChenTsao}.


An alternate, but equivalent, description of the process of loop formation for small $a$ can also be given.  After the endpoints are within a sphere of radius $b$, chain fluctuations will drive them in and out of the sphere many times before contact is established.  This allows us to describe the search process using an effective rate constant $\k_{eff}$, schematically shown in Fig. \ref{fig:kappa}.  For small $a$, the loop closure (a search within radius $b$) becomes effectively rate limited as opposed to diffusion limited \cite{Loop_expt_denature} contact formation.  The search will be successful, in the SSS formalism, on a time scale
\begin{eqnarray}
\t_{b\to a}\approx \frac{1}{2D_0\N'}\int_a^b \frac{dr}{P(r)}\blp\int_r^b dr'\ P(r')\brp^2,
\end{eqnarray}
with $\N'=\int_a^b dr\,P(r)$.  Again, we have taken $D=2D_0$ in this regime, because loop formation is dominated by the fast fluctuations of the monomers, which occur on the time scale of $b^2/D_0$.  For $a\approx b$, $\t_{b\to a}\approx (a-b)^2/6D_0$, whereas $\t_{b\to a}\approx b^3/6aD_0$ as $a\to 0$.  $\t_{b\to a}$ can be used to define the effective rate constant $\k_{eff}\propto (b-a)/\t_{b\to a}$.  This can be substituted into eq. (\ref{SSS}), and gives the approximate loop closure time as $a\to 0$
\begin{eqnarray}
\t_c(a)-\t_c(b)\approx \frac{1}{\k_{eff}\N'P(b)}\propto  \frac{N^{3/2}b^3}{D_0a},
\end{eqnarray}
reproducing the same scaling for small $a$ as in eq. (\ref{TauSmallA}).

The two-stage mechanism for the cyclization kinetics for $a/b<1$ is reminiscent of the two-state kinetic mechanism used to analyze experimental data.  The parameter $\k_{eff}$ is analogous to the reaction limited rate \cite{Loop_expt_denature}.  If the search rate within the capture region given by $\k_{eff}$ is small, then we expect the exponent $\a_\t<2$.  Indeed, the experiments of Buscaglia et. al. suggest that $\a_\t$ changes from 2 (diffusion-limited) to 1.65 (reaction-limited).  Our simulation results show the same behavior $\a_\t=2$ for $a/b\ge 1$, which corresponds to a diffusion limited process, and $\a_\t \approx 1.65$ for $a/b=0.1$, in which the search within $a/b<1$ becomes rate limiting.

\section{Loop Closure for Polymers in Good and Poor Solvents}

The kinetics of loop closure can change dramatically when interactions between monomers are taken into account. In good solvents, in which excluded volume interactions between the monomers dominate, it is suspected that only the scaling exponent in the dependence of $\tau_c$ on $N$ changes compared to Rouse chains. However, relatively little is known about the kinetics of loop closure in poor solvents in which enthalpic effects, that drive collapse of the chain, dominate over chain entropy. Because analytic work is difficult when monomer-monomer interactions become relevant we resort to simulations to provide insights into the loop closure dynamics.

\subsection{Simulation of Cyclization Times}

The Hamiltonian used in our simulations is $H=H_{FENE}+H_{LJ}$, where
\begin{eqnarray}
H_{FENE}=-\frac{kb^2}{2}\sum_{i=1}^N\log\bigg[1-\blp\frac{| \rv_{i+1}-\rv_{i}|-b}{R_0}\brp^2\bigg]
\end{eqnarray}
models the chain connectivity, with $k=22.2 k_B T$, and $b=0.38$nm.  The choice $R_0=2b/3$ (diverging at $|\rv_{i+1}-\rv_i|=b/3$ or $5b/3$) allowed for a larger timestep than using \cite{SOP} $R_0=b/2$, and increased the efficiency of conformational sampling.  The interactions between monomers are modeled using the Lennard-Jones potential,
\begin{eqnarray}
H_{LJ}=\e_{LJ}\,\sum_{i=1}^{N-2}\sum_{j=i+2}^N \bigg[\blp\frac{b}{\rv_{ij}}\brp^{12}-2\blp\frac{b}{\rv_{ij}}\brp^6\bigg],\label{HLJ}
\end{eqnarray}
with $\rv_{ij}=\rv_i-\rv_j$.  The Lennard-Jones interaction between the covalently bonded beads $\rv_{i}$ and $\rv_{i+1}$ are neglected to avoid excessive repulsive forces.  The second virial coefficient, defining the solvent quality, is given approximately by
\begin{eqnarray}\label{eq:virial2}
v_2(\e_{LJ})=\int d^3\rv \bigg[1-\exp\blp -\beta H_{LJ}(\rv)\brp\bigg],\label{virial_def}
\end{eqnarray}
with $\beta=1/k_B T$.  In a good solvent $v_2>0$, while in a poor solvent $v_2<0$.  A plot of $v_2$ as a function of $\e_{LJ}$ given in Fig. \ref{fig:virial}(A) shows that $v_2 >0$ when $\beta \e_{LJ}<0.3$ and $v_2 <0$ if $\beta \e_{LJ}>0.3$.  In what follows, we will refer to $\beta \e_{LJ}=0.4$ as weakly hydrophobic and $\beta \e_{LJ}=1.0$ as strongly hydrophobic. The classification of the solvent quality based on eq. (\ref{eq:virial2}) is approximate. The precise determination of the $\Theta$-point ($v_2\approx 0$) requiring the computation of $v_2$ for the entire chain. For our purposes, this approximate demarcation between good, $\Theta$, and poor solvents based on eq. (\ref{eq:virial2}) suffices.

To fully understand the effect of solvent quality on the cyclization time, we performed Brownian dynamics simulations for $\beta \e_{LJ}=i/10$, with $1\le i\le 10$.  In our simulations, $N$ was varied from 7 to 300 for each value of $\e_{LJ}$, with a fixed capture radius of $a=2b=0.76$nm.  The loop closure time was identified with the mean first passage time.  The dynamics for each trajectory was followed until the two ends were within the capture radius $a$.  Averaging the first passage times over 400 independent trajectories yielded the mean first passage time.  The chains were initially equilibrated using parallel tempering (replica exchange) Monte Carlo \cite{Frenkel} to ensure proper equilibration, with each replica pertaining to one value of $\e_{LJ}$.  In Fig. \ref{fig:virial}(B), we show the scaling of the radius of gyration $\la \Rv_{g}^2\ra$ as a function of $N$.  We find $\la\Rv^2_{ee}\ra\sim N^{\frac{6}{5}}$ for the good solvent and $\la\Rv^2_{ee}\ra\sim N$ for the $\Theta$ solvent ($\beta \e_{LJ}=0.3$).  In poor solvents ($\beta\epsilon_{LJ} >0.3$), the large $N$ scaling of $\la\Rv^2_{ee}\ra\sim N^{\frac{2}{3}}$ is not observed for the values of $N$ used in our simulations. Similar deviation from the expected scaling of $\la\Rv^2_{ee}\ra$ with $N$ have been observed by Rissanou et. al. \cite{Rissanou} for short chains in a poor solvent.  Simulations using much longer chains ($N \gtrsim 5000$) may be required to observe the expected scaling exponent of 2/3.


Brownian dynamics simulations with $D_0=0.77$nm$^2/$ns ($= k_B T/6\pi\eta b$, with $\eta=1.5$cP) were performed to determine $\t_c$.  The loop closure time for the chains in varying solvent conditions is shown in Fig. \ref{fig:tau_poor}(A) and (B).  The solvent quality drastically changes the loop closure time. The values of $\t_c$ for the good solvent ($\beta \e_{LJ}=0.1 $) are nearly three orders of magnitude larger than in the case of the strong hydrophobe ($\beta \e_{LJ}=1.0 $) for $N=80$ (Fig. \ref{fig:tau_poor}(A)).  For $N$ in the range of 20 to 30, that are typically used in experiments on tertiary contact formation in polypeptide chains, the value of $\tau_c$ is about 20ns in good solvents, whereas in poor solvents $\tau_c$ is only about 0.3ns. The results are vividly illustrated in Fig. \ref{fig:tau_poor}(B), which shows $\tau_c$ as a function of $\epsilon_{LJ}$ for various $N$ values. The differences in $\tau_c$ are less pronounced as $N$ decreases (Fig. \ref{fig:tau_poor}(B)).  The absolute value of $\t_c$ for $N\approx 20$ is an order of magnitude less than obtained for $\t_c$ in polypeptides~\cite{Loop_expt_denature}. There could be two inter-related reasons for this discrepancy. The value of $D_0$, an effective diffusion constant in the SSS theory, extracted from experimental data and simulated $P(\Rv_{ee})$ is about an order of magnitude less than the $D_0$ in our paper. Secondly, Buscaglia et al.~\cite{Loop_expt_denature} used the WLC model with excluded volume interactions whereas our model does not take into account the effect of bending rigidity. Indeed, we had shown in an earlier study~\cite{CBDT} that chain stiffness increases $\tau_c$. Despite these reservations, our values of $\tau_c$ can be made to agree better with experiments using $\eta \approx 5$cP~\cite{ToanPRL06} and a slightly larger value of $b$. Because it is not our purpose to quantitatively analyze cyclization kinetics in polypeptide chains we did not perform such comparison.


We also find that the solvent quality significantly changes the scaling of $\t_c\sim N^{\a_\t}$, as shown in Fig. \ref{fig:tau_poor}(C). For the range of $N$ considered in our simulations, $\t_c$ does not appear to vary as a simple power law in $N$ (much like $\la\Rv_g^2\ra$; see Fig. \ref{fig:virial}(B)) for $\beta\e_{LJ}>0.3$.  The values of $\tau_c$ in poor solvents shows increasing curvature as $N$ increases.  However, if we insist that a simple power law describes the data then for the smaller range of $N$ from 7 to 32 (consistent with the methods of other authors \cite{poor_expt_2,Loop_expt_denature,poor_expt_1}), we can fit the initial slopes of the curves to determine an effective exponent $\a_\t$ (\ref{fig:tau_poor}(C)), i.e. $\t_c\approx \t_0 N^{\a_\t}$.  In the absence of sound analytical theory, the extracted values of $\alpha_\tau$ should be viewed as an effective exponent.  We anticipate that, much like the scaling laws for $\la\Rv_g^2\ra$, the final large $N$ scaling exponent for $\t_c$ will only emerge for \cite{Rissanou} $N\gtrsim 5000$, which is too large for accurate simulations.  However, with the assumption of a simple power law behavior for small $N$, we find that the scaling exponent precipitously drops from $\a_\t\approx 2.4$ in the good solvent to $\a_\t\approx 1.0$ in the poor solvent.  Our estimate of $\a_\t$ in good solvents is in agreement with the prediction of Debnath and Cherayil \cite{Debnath} ($\a_\t\approx 2.3-2.4$) or Thirumalai \cite{Thirum_JPCB} ($\a_\t\approx 2.4$), and is fairly close to the value obtained in previous simulations\cite{Alexi} ($\a_\t\approx 2.2$).  The differences between the simulations may be related to the choice of the Hamiltonian.  Podtelezhnikov and Vologodskii \cite{Alexi} used a harmonic repulsion between monomers to represent the impenetrability of the chain, and took $a/b<1$ in their simulations.

In contrast to the good solvent case, our estimate of $\a_\t$ in poor solvents is significantly lower than the predictions of Debnath and Cherayil \cite{Debnath},  who suggested $\a_\t\approx 1.6-1.7$, based on a modification of the WF formalism \cite{WF}.  However, fluorescence experiments on multiple repeats of the possibly weakly hydrophobic glycine and serine residues in D$_2$O have found $\t_c\sim N^{1.36}$ for short chains \cite{poor_expt_2} and $\t_c\sim N^{1.05}$ for longer chains \cite{poor_expt_1}, in qualitative agreement with our simulation results.  Bending stiffness \cite{Bovine,CBDT} and hydrodynamic interactions may make direct comparison between these experiments and our results difficult.  The qualitative agreement between simulations and experiments on polypeptide chains suggest that interactions between monomers are more important than hydrodynamic interactions, which are screened.

\subsection{Mechanisms of Loop Closure in Poor Solvents}

The dramatically smaller loop closure times in poor solvents than in good solvents (especially for $N > 20$; see Fig \ref{fig:tau_poor}(B)) requires an explanation. In poor solvents, the chain adopts a globular conformation with the monomer density $\rho b^3 \sim {\cal O}(1)$, where $\rho \approx N/R_g^3$. We expect the motions of the monomers to be suppressed in the dense, compact globule.  For large $N$, when entanglement effects may dominate, it could be argued that in order for the initially spatially separated chain ends ($|{ \Rv}_{ee}|/a >1$) to meet, contacts between the monomer ends with their neighbors must be broken. Such unfavorable events might require overcoming enthalpic barriers ($\approx \bar Q\times \epsilon_{LJ}$, where $\bar Q$ is the average number of contacts for a bead in the interior of the globule), which would increase $\tau_c$.  Alternatively, if the ends search for each other using a diffusive, reptation-like mechanism without having to dramatically alter the global shape of the collapsed globule, $\tau_c$ might decrease as $\epsilon_{LJ}$ increases (i.e. as the globule becomes more compact). It is then of interest to ask whether looping events are preceded by global conformational changes, with a large scale expansion of the polymer that allows the endpoints to search the volume more freely, or if the endpoints search for each other in a highly compact, but more restrictive, ensemble of conformations.


In order to understand the mechanism of looping in poor solvents, we analyze in detail the end-to-end distance $|\Rv_{ee}(t)|$ and the radius of gyration $|\Rv_g(t)|$ for two trajectories (with $\beta\e_{LJ}=1$ and $N=100$). One of the trajectories has  a fast looping time (${\t_{c}}_F\approx 0.003$ns), while the looping time in the other is considerably slower (${\t_{c}}_S\approx 4.75$ns).  Additionally, we compute the time-dependent variations of the coordination number, $Q(t)$ for each endpoint.  We define two monomers $i$ and $j$ to be in `contact' if $|\rv_{i}-\rv_j|\ge 1.23 b$ (beyond which the interaction energy $E_{LJ}\ge -\e_{LJ}/2$), and define $Q_1(t)$ and $Q_N(t)$ to be the total number of monomers in contact with monomers 1 and $N$ respectively.  We do not include nearest neighbors on the backbone when computing the coordination number, and the geometrical constraints gives $0\le Q(t)\le 11$ for either endpoint.  With this definition, an endpoint on the surface of the globule will have $Q=5$.  These quantities are shown in Figs \ref{fig:Q_plot} and \ref{fig:Q_plot2}.

The trajectory with ${\tau_c}_F$ (Fig \ref{fig:Q_plot}) shows little variation in either $|\Rv_g|$ or $|\Rv_{ee}|$.  We find $|\Rv_{ee}|\approx |\Rv_g|$, suggesting that the endpoints remain confined within the dense globular structure throughout the looping process.  This is also reflected in the coordination numbers for both of the endpoints, with both $Q_1(t)$ and $Q_N(t)$ are in the range $5\le Q(t)\le 10$ throughout the simulation.  The endpoints in this trajectory, with the small loop closure time $\t_c^F$, always have a significant number of contacts, and traverse the interior of the globule when searching for each other.  Similarly, we also found that the trajectory with a long first passage time $\t_{c}^S$ (Fig \ref{fig:Q_plot2}) shows little variation in $\Rv_g$ throughout the run.  The end-to-end distance, however, shows large fluctuations over time, and $\la \Rv_{ee}^2\ra\gtrsim 2\la\Rv_g^2\ra$ until closure.  This suggests that, while the chain is in an overall globular conformation (small, constant $\Rv_g^2$), the endpoints are mainly found on the exterior of the globule.  This conclusion is again supported by the coordination number, with $Q(t)\le 5$ for significant portions of the simulation.  While the endpoints are less restricted by nearby contacts and able to fluctuate more, the endpoints spend much longer time searching for each other.  Thus, it appears that the process of loop formation in poor solvents, where enthalpic effects might be expected to dominate for $N=100$ occurs by a diffusive, reptation-like process. Entanglement effects are not significant in our simulations.

We note that trajectories in which the first passage time for looping is rapid (with ${\t_c}_i<\t_c$ for trajectory $i$) have at least one endpoint with a high coordination number ($Q>5$) throughout the simulation. In contrast for most slow-looping run (with ${\t_c}_i>\t_c$), we observe long stretches of time where both endpoints have a low coordination number ($Q<5$).  These results suggest that motions within the globule are far less restricted than one might have thought, and loop formation will occur faster when the endpoints are within the globule than it would if the endpoints were on the surface. The longer values of $\tau_c$ are found if the initial separation of the end points is large, which is more likely if they are on the surface than buried in the interior.  The absence of any change in $|\Rv_g(t)|$ in both the trajectories, which represent the extreme limits in the first passage time for looping, clearly shows that contact formation in the globular phase is not an activated process.  Thus, we surmise that looping in poor solvents occurs by a diffusive, reptation-like mechanism, provided entanglement effects are negligible.


\subsection{Separating the Equilibrium Distribution $P(\Rv_{ee})$ and Diffusive Processes in Looping Dynamics}

The results in the previous section suggest a very general mechanism of loop closure for interacting chains.  The process of contact formation for a given trajectory depends on the initial separation $\Rv_{ee}$, and the dynamics of the approach of the ends.  Thus, $\t_c$ should be determined by the distribution of $P(\Rv_{ee})$ (an equilibrium property), and an effective diffusion coefficient $D(t)$ (a dynamic property).  We have shown for the Rouse model that such a deconvolution into an equilibrium and dynamic part, which is in the spirit of the SSS approximation, is accurate in obtaining $\t_c$ for a wide range of $N$ and $a/b$.  It turns out that a similar approach is applicable to interacting chains as well.

The decomposition of looping mechanisms into a convolution of equilibrium and dynamical parts explains the large differences in $\t_c$ as the solvent quality changes.  We find, in fact, that the equilibrium behavior of the endpoints dominates the process of loop formation, with the kinetic processes being only weakly dependent on the solvent quality for short chains.  In Fig. \ref{fig:pdf}(A), we plot the end-to-end distribution function for weakly ($\beta\e_{LJ}=0.4$) and strongly ($\beta\e_{LJ}=1$) hydrophobic polymer chains.  The strongly hydrophobic chain is highly compact, with a sharply peaked distribution.  The average end-to-end distance is significantly lower than is the weakly hydrophobic case.  While the distribution function is clearly strongly dependent on the interactions, the diffusion coefficient $D(t)$ is only weakly dependent on the solvent quality (Fig. \ref{fig:pdf}(B)).  The values of $D(t)=\la\d\Rv_{ee}^2\ra/6t$ are only reduced by a factor of about 2 between the $\beta\e_{LJ}=0.1$ (good solvent, with a globally swollen configuration) and the $\beta\e_{LJ}=1.0 $ (poor solvent, with a globally globular configuration) on intermediate time scales.  We note, in fact, that the good solvent and $\Theta$ solvent cases have virtually identical diffusion coefficients throughout the simulations (Fig \ref{fig:pdf}(B)).  This suggests that the increase in $\t_c$ (Fig \ref{fig:tau_poor}) between the Rouse chains and the good solvent chains is primarily due to the broadening of the distribution $P(\Rv_{ee}$), i.e. the significant increase in the average end-to-end distance in the good solvent case, $\la \Rv_{ee}^2\ra\sim N^{2\nu}$, with $\nu=3/5$.


Because of the weak dependence of the diffusion coefficient on the solvent quality, the loop closure time is dominated primarily by the end-to-end distribution function.   In other words, the equilibrium distribution function $P(\Rv_{ee})$, to a large extent, determines $\t_c$.  To further illustrate these arguments, we find that if we take $D\approx 2D_0$ in eq. (\ref{SSS}) and numerically integrate the distribution function found in the simulations for $N=100$, $\t_c(\beta\e_{LJ}=1.0)$ and $\t_c(\beta\e_{LJ}=0.4)$ differ by two orders of magnitude, almost completely accounting for the large differences seen in Fig.  \ref{fig:tau_poor}(B) between the two cases.  Moreover, if the numerically computed values of $D(t)$ for long $t$ ($t > 0.5$ns in Fig.~\ref{fig:pdf} for example) is used for $D_{ee}$ in eq. (\ref{SSS}) we obtain values of $\t_c$ that are in reasonably good agreement with simulations. The use of $D_{ee}$ ensures that the dynamics of the entire chain is explicitly taken into account. These observations rationalize the use of $P(\Rv_{ee})$ with a suitable choice of $D_{ee}$ in obtaining accurate results for flexible as well as stiff chains \cite{CBDT,JunHa}. Because $P(\Rv_{ee})$ can, in principle, be inferred from FRET experiments~\cite{FRET1,FRET2} the theory outlined here can be used to quantitatively predict loop formation times. In addition, FRET experiments can also be used to assess the utility of polymer models in describing fluctuations in single stranded nucleic acids and polypeptide chains.

\subsection{Kinetics of interior loop formation}
We computed the kinetics of contact between beads that are in the chain interior as a function of solvent quality (Fig.~\ref{fig:rl}(A)) using $N = 32$. The mean time for making a contact is computed using the same procedure as used for cyclization kinetics. For simplicity we only consider interior points that are equidistant (along the chain contour) from the chain ends. The ratio $r_l$, which measures the change in the time for interior loop formation relative to cyclization kinetics, depends on $\beta\epsilon_{LJ}$ \textit{and} $l/N$, where $l$ is the separation between the beads (Fig.~\ref{fig:rl}(A)). The non-monotonic dependence of $r_l$ on $l$ in good solvents further shows that as $l/N$ decreases to about $0.6$, $r_l\approx 1$. The maximum in $r_l$ at $l/N\approx 0.9$ decreases as $\beta\epsilon_{LJ}$ increases. In the poorest solvents considered ($\beta\epsilon_{LJ} = 0.8$), we observe that $r_l$ only decreases monotonically with decreasing $l/N$. Interestingly, in poor solvents, $r_l$ can be much less than unity which implies that it is easier to establish contacts between beads in the chain interior than between the ends. This prediction can be verified in polypeptide chains in the presence of inert crowding agents that should decrease the solvent quality. Just as in cyclization kinetics, interior loop formation also depends on the interplay between internal chain diffusion that gets slower as the solvent quality decreases and equilibrium distribution (which gets narrower) of the distance between the contacting beads.

We also performed simulations for $N=80$ by first computing the time for cyclization $\tau_c^{80}$. In another set of simulations, two flexible linkers each containing $20$ beads were attached to the ends of the $N = 80$ chain. For the resulting longer chain we calculated $\tau_l$ for $l=80$ as a function of $\beta\epsilon_{LJ}$. Such a calculation is relevant in the context of single molecule experiments in which the properties of a biomolecule (RNA) is inferred by attaching linkers with varying polymer characteristics. It is important to choose the linker characteristics that minimally affects the dynamic properties of the molecule of interest.  The ratio $\tau_{l=80}/\tau_c^{80}$ depends on $\beta\epsilon_{LJ}$ and changes from $2.6$ (good solvents) to $2.0$ under $\Theta$ condition and becomes unity in poor solvents (Fig.~\ref{fig:rl}(B)). Analysis on the dependence of the diffusion coefficients of interior-to-interior vector $D_{ij}$ ($i=20$ and $j = 100$) and end-to-end vector (of original chain without linkers) $D_{ee}$ on solvent conditions indicates that on the time scales relevant to loop closure time (analogous to $\tau_{ee}$ for the Rouse chain), $D_{ij}$ reduces to about one half of $D_{ee}$ in good and $\Theta$ solvents, whereas the two are very similar in poor solvents. The changes in the diffusion coefficient together with the equilibrium distance distribution explains the behavior in Fig.~\ref{fig:rl}(B).

\section{Conclusions}

A theoretical description of contact formation between the chain endpoints is difficult because of the many body nature of the dynamics of a polymer.  Even for the simple case of cyclization kinetics in Rouse chains, accurate results for $\t_c$ are difficult to obtain for all values of $N$, $a$, and $b$.  The present work confirms that, for large $N$ and $a/b>1$, the looping time must scale as $N^2$, a result that was obtained some time ago using the WF formalism \cite{WF,Doi_Chem_Phys}.  Here, we have derived $\t_c\sim N^2$ (for $N\gg 1$ and $a\ge b$) by including the full internal chain dynamics within the simple and elegant SSS theory \cite{SSS}.  We have shown that, for $N<100$ and especially in the (unphysical) limit $a/b<1$, the loop closure time $\t_c\sim \t_0 N^{\a_\t}$ with $1.5<{\a_\t}\le 2$.  In this limit, our simulations show that loop closure occurs in two stages with vastly differing time scales.  By incorporating these processes into a scale-dependent diffusion coefficient, we obtain an expression for $\t_c$ that accurately fits the simulation data.  The resulting expression for $\t_c$ for $a<b$ (eq. (\ref{TauSmallA})) contains both the $N^{\frac{3}{2}}$ and $N^2$ limits, as was suggested by Pastor et. al. \cite{PastorZwanzig}

The values of $\t_c$ for all $N$ change dramatically when interactions between monomers are taken into account.  In good solvents, $\t_c\sim \t_0 N^{\a_\t}$ ($\a_\t\approx 2.4$) in the range of $N$ used in the simulations.  Our exponent $\a_\t$ is in reasonable agreement with earlier theoretical estimates \cite{Thirum_JPCB,Debnath}.  Polypeptide chains in high denaturant concentrations may be modeled as flexible chains in good solvents.  From this perspective, the simple scaling law can be used to fit the experimental data on loop formation in the presence of denaturants using physical values of $\t_0$.  Only when $N$ is relatively small ($N\approx 4$) will chain stiffness play a role in controlling loop closure times.  Indeed, experiments show that $\t_c$ increases for short $N$ (see Fig. 3 in Ref. 15), and deviates from the power law behavior given in eq. (\ref{SSSOld}) for all $N$, which is surely due to the importance of bending rigidity.

The simulation results for $\t_c$ in poor solvents show rich behavior that reflects the extent to which the quality of the solvent is poor.  The poorness of the solvent can be expressed in terms of
\begin{eqnarray}
\lambda=\frac{\e_{LJ}-\e_{LJ}(\Theta)}{\e_{LJ}(\Theta)}
\end{eqnarray}
where the $\Theta$-solvent interaction strength $\beta\e_{LJ}(\Theta)\approx 0.3 $ is determined from $v_2\approx 0$ (Fig. \ref{fig:virial}).  Loop closure times decrease dramatically as $\lambda$ increases.  For example, $\t_c$ decreases by a factor of about 100 for $N=80$ as $\lambda$ increases from 0 to 2.3.  In this range of $N$, a power law fit of $\t_c$ with $N$ ($\t_c\sim N^{\a_\t}$) shows that the exponent $\a_\t$ depends of $\lambda$.  Analysis of the trajectories that monitor loop closure shows that contact between each end of the chains is established by mutual, reptation-like motion within the dense, compact globular phase.

The large variations of $\t_c$ as $\lambda$ changes suggests that there ought to be significant dependence of loop formation rates on the sequence in polypeptide chains.  In particular, our results suggest that as the number of hydrophobic residues increase, $\t_c$ should decrease.  Similarly, as the number of charged or polar residues increase, the effective persistence length ($l_p$) and interactions can be altered, which in turn could increase $\t_c$.  Larger variations in $\t_c$, due to its dependence on $l_p$ and $N$, can be achieved most easily in single-stranded RNA and DNA.  These arguments neglect sequence effects, which are also likely to be important. The results in Fig.~\ref{fig:tau_poor}(B) may also be reminiscent of ``hydrophobic collapse'' in proteins especially as $\lambda$ becomes large. For large $\lambda$ and long $N$ it is likely that $\tau_c$ correlates well with time scales for collapse. This scenario is already reflected in $P(\Rv_{ee})$ (see Fig.~\ref{fig:pdf}(A)). It may be possible to discern the predictions in Fig.~\ref{fig:tau_poor}(B) by varying the solvent quality for polypeptides. Combination of denaturants (makes the solvent quality good) and PEG (makes it poor) can be used to measured $\tau_c$ in polypeptide chains. We expect the measured $\tau_c$ should be qualitatively similar to the findings in Fig.~\ref{fig:tau_poor}(B).

The physics of loop closure for small and intermediate chain lengths ($N \le 300$) is rather complicated, due to contributions from various time and length scales (global relaxation and internal motions of the chains). The contributions from these sources are often comparable, making the process of looping dynamics difficult to describe theoretically. A clear picture of the physics is obtained only when one considers all possible ranges of the parameters entering the loop closure time equation. To this end, we have explored wide ranges of conceivable parameters, namely the chain length $N$, capture radius $a$, and conditions of the solvents expressed in terms of $\epsilon_{LJ}$.  By combining analytic theory and simulations, we have shown that, for a given $N$, the looping dynamics in all solvent conditions is primarily determined by the initial separation of the end points.  The many body nature of the diffusive process is embodied in $D(t)$, which does not vary significantly as $\lambda$ changes for a fixed $N$.  Finally, the dramatic change in $\t_c$ as $\lambda$ increases suggests that it may be also necessary to include hydrodynamic interactions, that may decrease $\t_c$ further, to more accurately obtain the loop closure times.


\section{Appendix A}

Friedman and O'Shaughnessy \cite{FriedmanJPII91} (FO) generalized the concept of \emph{the exploration of space}
suggested by de Gennes \cite{deGennesJCP82} to the cyclization reaction of
polymer chain.
The arguments given by de Gennes and FO succinctly reveal the conditions under which local equilibrium is appropriate in terms of properties of the polymer chains.

First, de Gennes introduced the notion of compact and non-compact exploration of space
associated with a bimolecular reaction involving polymers.  Tertiary contact formation is a particular example of such a process.
Consider the relative position between two reactants on a lattice with the lattice spacing $a$.
The two reactants explore the available conformational space until their relative distance becomes less than the reaction radius.
One can define two quantities relevant to the volume spanned prior to the reaction.
One comes from the actual number of jumps on the lattice defined as $j(t)$
which is directly proportional to $t$.
If the jump is performed in a d-dimensional lattice, the actual volume explored would be $a^dj(t)$.
The other quantity comes from the root mean square distance.
If $x(t)\sim t^u$ is the root mean square distance for one-dimension, $x^d(t)$ is the net
volume explored.
The comparison between these two volumes defines the compactness in the exploration of the space.
i)  The case $x^d(t)>a^dj(t)$ corresponds to non-compact exploration of the space ($ud>1$).
ii) The regime $x^d(t)<a^dj(t)$ represents compact exploration of the space ($ud<1$).
Depending on the dimensionality, the exploration of space by the reactive pair in
the bimolecular reaction is categorized either into non-compact (d=3) or into
compact (d=1) exploration.
In case of non-compact exploration, the bimolecular reaction takes place infrequently,
so that the local equilibrium in solution is easily reached.
The reaction rate is simply proportional to
the probability that the reactive pair is within the reaction radius,
so that $k\sim p_{eq}(r<r_0)$, which eventually
leads to $k=4\pi\sigma D$, the well-known steady state diffusion controlled
rate coefficient.
It can be shown that $k\sim t^{ud-1}$ in case of compact exploration.

In the context of polymer cyclization the compactness of the exploration of space can be assessed using the exponent $\theta=\frac{d+g}{z}$, where $g$ is the correlation hole exponent, $z$ is the dynamic exponent.
Since \cite{deGennesbook, desCloizeauxJP80} $\lim_{r\rightarrow 0}p_{eq}(r)\sim\frac{1}{R^d}\left(\frac{r}{R}\right)^g$
and the cyclization rate can be approximated by $k\sim\frac{d}{dt}\int d^dr\, p_{eq}(r)$, it follows that
$k\sim\frac{d}{dt}\left(\frac{r}{R}\right)^{d+g}$. The relations $r\sim t^{1/z}$ and $R\sim\tau^{1/z}$
lead to $k\sim\frac{1}{\tau}\left(\frac{t}{\tau}\right)^{\frac{d+g}{z}-1}$, where $\t$ is the characteristic relaxation time.
\begin{enumerate}
\item
If $\theta>1$, then the cyclization rate is
given by $k\sim p_{eq}(r=r_0)\sim\frac{1}{R^d}f(\frac{r}{R})$.
which, with $R\sim N^{\nu}$, leads to the scaling relation
\begin{equation}
\tau_c\sim N^{\nu(d+g)}.\label{A1}
\end{equation}
\item
If $\theta<1$, the compact exploration of conformations occurs between the chain ends.
As a result, the internal modes are not in local equilibrium.
In this case, $\tau_c\sim\tau_R\sim R^z$ where $z=2+\frac{1}{\nu}$ is the dynamic exponent
for free draining case and $z=d$ when hydrodynamic interactions are included \cite{deGennesbook, EdwardsBook}.
Therefore, the scaling law for cyclization rate is given by
\begin{equation}
\tau_c\sim N^{z\nu}.
\label{eqn:scaling2}
\end{equation}
\end{enumerate}
The inference about the validity of local equilibrium, based on $\theta$, is extremely useful in obtaining the scaling laws for polymer
cyclization, eqs. (\ref{A1}) and (\ref{eqn:scaling2}).  Extensive Brownian dynamics simulation by Rey et.al.\cite{ReyMacro98_2} have established the validity
of these scaling laws.
The expected scaling laws for three different polymer models are discussed below.
\begin{itemize}
\item
Free-draining Gaussian chain ($d=3$, $g=0$, $z=4$, $\nu=\frac{1}{2}$) :
$\theta=\frac{3}{4}<1$.

Because $\theta<1$, the local equilibrium approximation
is not valid for a ``long'' free-draining Gaussian chain, or equivalently the Rouse model.
Accordingly, we expect $\tau_c\sim N^2$ for the Rouse chain for $N\gg 1$.
However, if $N$ is small and the local equilibrium is established among
the internal Rouse modes so that $\tau_c\gg\tau_R$,
the scaling relation change from $\tau_c\sim N^2$ to  $\tau_c\sim N^{\a_\t}$, with ${\a_\t}<2$.
The simulations shown here and elsewhere, \cite{PastorZwanzig} and the theory by Sokolov, \cite{Sokolov} explicitly demonstrates that ${\a_\t}$ can be less than 2 for small $N$.
In this sense looping time of free-draining Gaussian chain of finite size is bound
by \cite{Portman,CBDT} $\tau_{SSS}<\tau_c<\tau_{WF}$.
\item
Free-draining Gaussian chain with excluded volume ($d=3$, $g=\frac{\gamma-1}{\nu}=\frac{5}{18}$, $z=\frac{11}{3}$, $\nu=\frac{3}{5}$) : $\theta=\frac{59}{66}<1$.

From eq. (\ref{eqn:scaling2}), it follows that $\tau_c\sim N^{2.2}$
This polymer model has been extensively studied using Brownian dynamaics simulation and the value of the scaling exponent 2.2 has been confirmed by Vologodskii \cite{Alexi}.  The value of the exponent (2.2) is also consistent with previous theoretical predictions \cite{Debnath,Thirum_JPCB}.
\item
Gaussian chain with excluded volume and hydrodynamic interactions
($d=3$, $g=\frac{5}{18}$, $z=3$, $\nu=\frac{3}{5}$) : $\theta=\frac{59}{54}>1$.

Since $\theta>1$, the local equilibrium approximation is expected to hold.
This polymer model corresponds to the flexible polymer in a good solvent.
The incorporation of hydrodynamic interactions may assist the fast relaxation of the rapid internal modes,
and changes the nature of cyclization dynamics from a compact to a non-compact one.
The correct scaling law is predicted to be $\tau_c\sim N^{2.0}$.
Since the local equilibrium approximation is correct,
the first passage time approach \cite{SSS} should give a correct estimate of $\tau_c$
only if the effective potential of mean force acting on the two ends of the chain is known.
\end{itemize}

\section{Appendix B}
The relation between the mean first passage time $\tau$ and the probability $\Sigma(t)$ that at time $t$ the system is still unreacted is {\em exact}:
\begin{eqnarray}\label{eq:tau_sigmas}
 \tau &=& \int_0^\infty \Sigma(t) dt,
\end{eqnarray}
for any form of $\Sigma(t)$ for which $\Sigma(0)$ is finite and $\lim_{t\rightarrow\infty} t\Sigma(t) = 0$. Therefore, the stricter requirement that $\Sigma(t) \sim \exp(-t/\tau)$ in the SSS original paper~\cite{SSS} is not necessary.

We define $F(t)$ the \emph{flux} (or {\em density}) {\em of passage}: $F(t) \equiv -{\partial \Sigma(t)\over \partial t}$. The mean first passage time is:
\begin{eqnarray}\label{eq:tau_St}
 \tau =\int_{0}^\infty t F(t) dt=\int_{0}^\infty t \left( -{\partial \Sigma(t)\over \partial t}\right)dt = - \int_{0}^\infty t\ d\Sigma(t).
\end{eqnarray}
Performing integration by parts gives:
\begin{eqnarray}\label{eq:tau_sigma1}
 \tau &=& -\left. {t\Sigma(t)}\right\vert_0^\infty +\int_0^\infty \Sigma(t) dt.
\end{eqnarray}
By definition, $\Sigma(t)$ must be finite, and hence $t\Sigma(t) = 0$ at $t=0$. If $\Sigma(t)$ is such that it vanishes at $t\rightarrow \infty$ faster than $t^{-1}$, then the first term in eq. (\ref{eq:tau_sigma1}) vanishes and we are left with eq.~(\ref{eq:tau_sigmas}). Note that these are also necessary and sufficient conditions for $\tau$ in eq.~(\ref{eq:tau_sigmas}) to be finite.

\section{Appendix C}

In formulating the fluctuations of the end-to-end distance vector, $\la\d\Rv_{ee}^2\ra$, it is important to take into account the failings of the continuum model of the Freely Jointed Chain.  A simple calculation of $\la \d\Rv_{ee}^2(t)\ra$ with $\Rv_{ee}(t)=\rv(N,t)-\rv(0,t)$ as determined from eq (\ref{Langevin}) gives
\begin{eqnarray}
\la \d\Rv_{ee}^2(t)\ra=16Nb^2\sum_{n\mbox{ odd}}\frac{1}{n^2\pi^2}\blp1-e^{-n^2t/\t_R}\brp\label{standard}
\end{eqnarray}
We will refer to this result as the standard analytic average.  However, the non-physical boundary conditions imposed on the continuum representation, with $\partial\rv/\partial s\equiv0$ at the endpoints, will strongly affect the accuracy of this result.

To minimize the effect of the boundary conditions on averages involving the end-to-end distance, we compute averages with respect to the differences between the centers of mass of the first and last bonds, using
\begin{eqnarray}
\Rv_{ee}(t)\approx \int_{N-1}^N ds\ \rv(s,t)-\int_0^1 ds\  \rv(s,t).
\end{eqnarray}
We will refer to this as the center of mass average.  Using this representation, $\la \d\Rv_{ee}^2(t)\ra$ is given in eq (\ref{Ree}).

In Fig. \ref{fig:diff}, we compare the values of $D(t)$ obtained from $\la \d\Rv_{ee}^2(t)\ra$ (in eq. (\ref{dt})) for $N=19$ and $b=0.39$.  In both cases, $b$ is taken as a fitting parameter.  The center of mass average, which fits the data quite well, has a best fit of $b=0.41$ (a difference of 5\%), whereas the standard average does not give accurate results.  For this reason, all averages involving $\Rv_{ee}$ are computed using the center of mass theory.

{\bf Acknowledgments}:  This work was supported in part by a grant from the National Science Foundation through grant number NSF CHE 05-14056.

\eject

\singlespacing


\providecommand{\refin}[1]{\\ \textbf{Referenced in:} #1}

\eject

{\bf{Figure Captions}}\\

\begin{figure}[htbp] 
   \centering

      \includegraphics[width=.8\textwidth,angle=0]{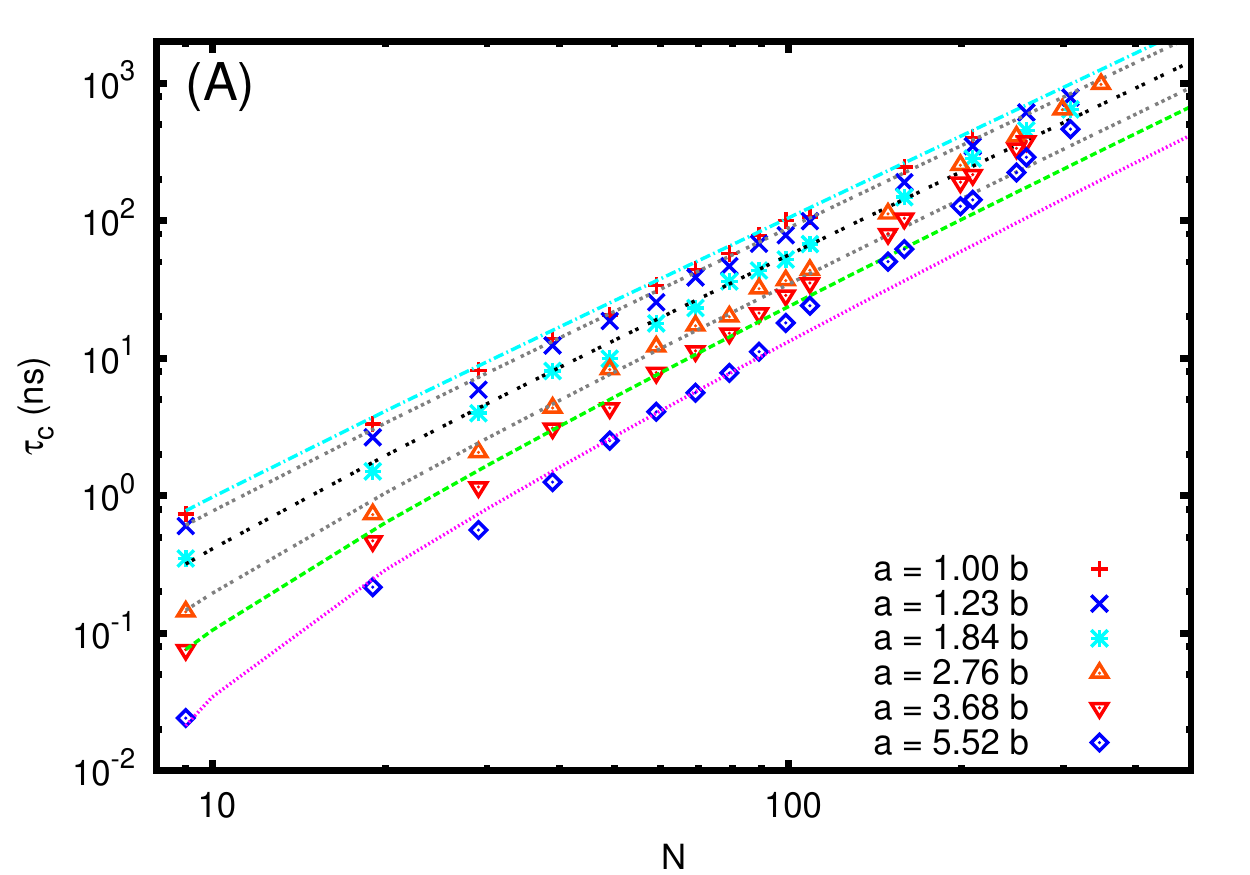} \\
            \includegraphics[width=.8\textwidth,angle=0]{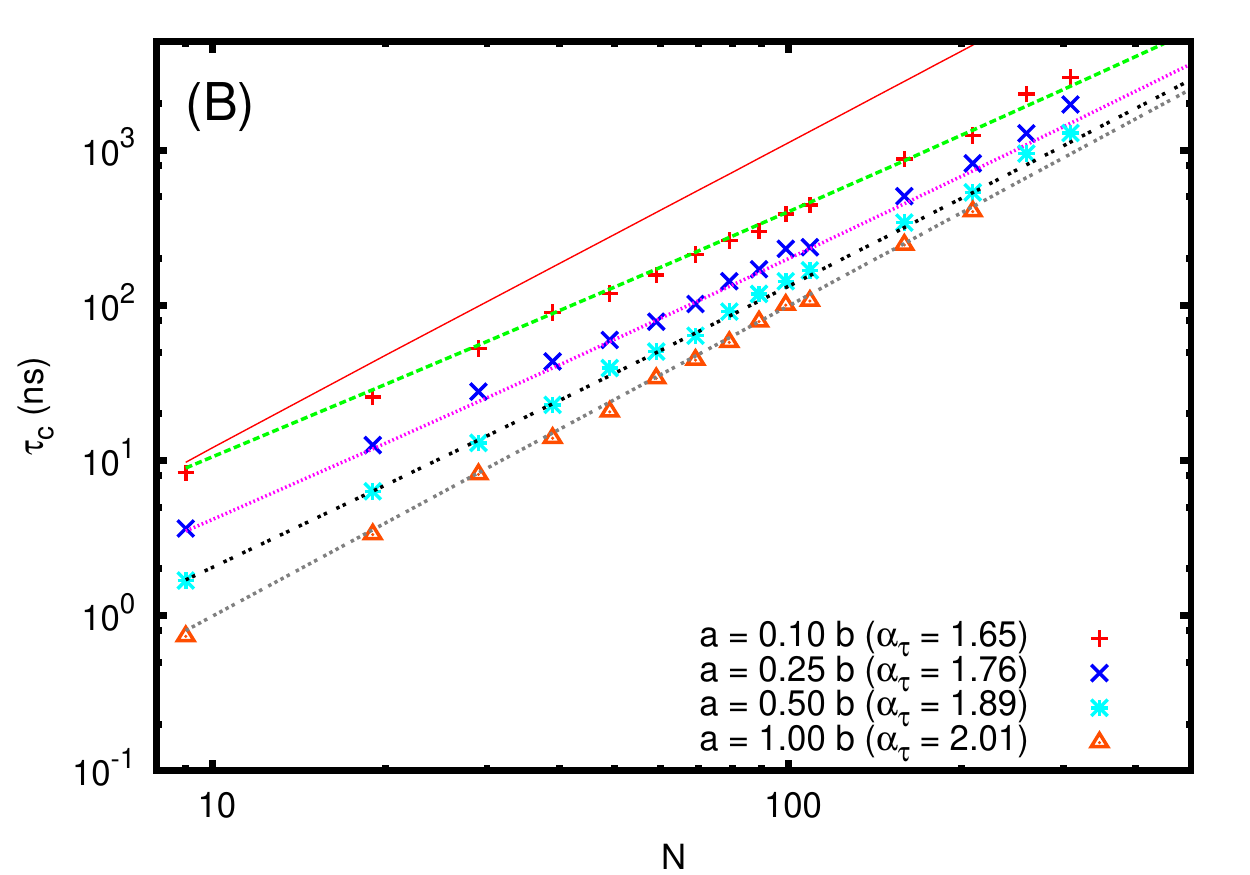}

   \caption{}
      \label{fig1}
\end{figure}

{\underline{Fig. \ref{fig1}}}:  Dependence of $\t_c$ on $N$ for various values of $a$.  The symbols correspond to different values of the capture radius.  (A):  The values of $a/b$ are  1.00 (+), 1.23 ($\times$), 1.84 ($*$), 2.76 ($\Delta$), 3.68 ($\nabla$), and 5.52 ($\Diamond$).  The lines are obtained using eq. (\ref{SSS}) with $\k\to\infty$.  The diffusion constant in eq. (\ref{SSS}) is obtained using $D=\la\d\Rv^2_{ee}(\t_{ee}/2)/3\t_{ee}$, with $\la\d\Rv_{ee}^2(t)\ra$ given in eq. (\ref{Rfluct}).  (B):  The values of $a/b$ are 0.10 (+), 0.25 ($\times$), 0.50 ($*$), and 1.00 ($\Delta$).  The lines are the theoretical predictions using eq. (\ref{TauSmallA}). The poor fit using eq. (\ref{SSS_new}) with $a=0.1 b$ (solid line) shows that the two-stage mechanism has to be included to obtain accurate values of $\tau_c$.  The effective exponent $\a_\t$, obtained by fitting $\t_c\sim N^{\a_\t}$, is shown in parentheses.
\\

\begin{figure}[htbp] 
   \centering
   \includegraphics[width=.8\textwidth]{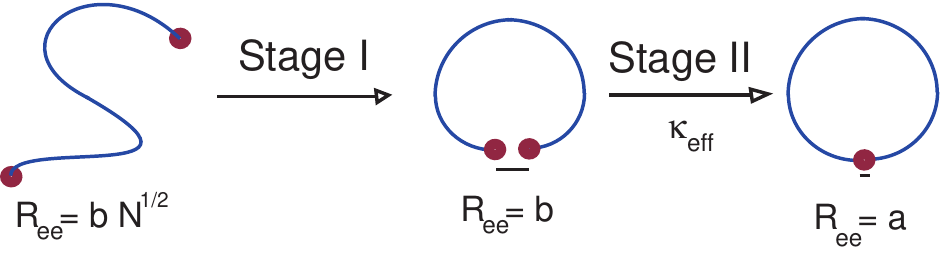}
   \caption{}
   \label{fig:kappa}
\end{figure}

{\underline{Fig. \ref{fig:kappa}}}:  Sketch of the two-stage mechanism for loop closure for Rouse chains when $a<b$.  Although unphysical, this case is of theoretical interest.  In the first stage, fluctuations in $\Rv_{ee}$ result in the ends approaching $|\Rv_{ee}|=b$.  The search of the monomers within a volume $b^3$ ($>a^3)$, which is rate limiting, leads to a contact in the second stage.
\\

\begin{figure}[htbp] 
   \centering
   \includegraphics[width=.8\textwidth]{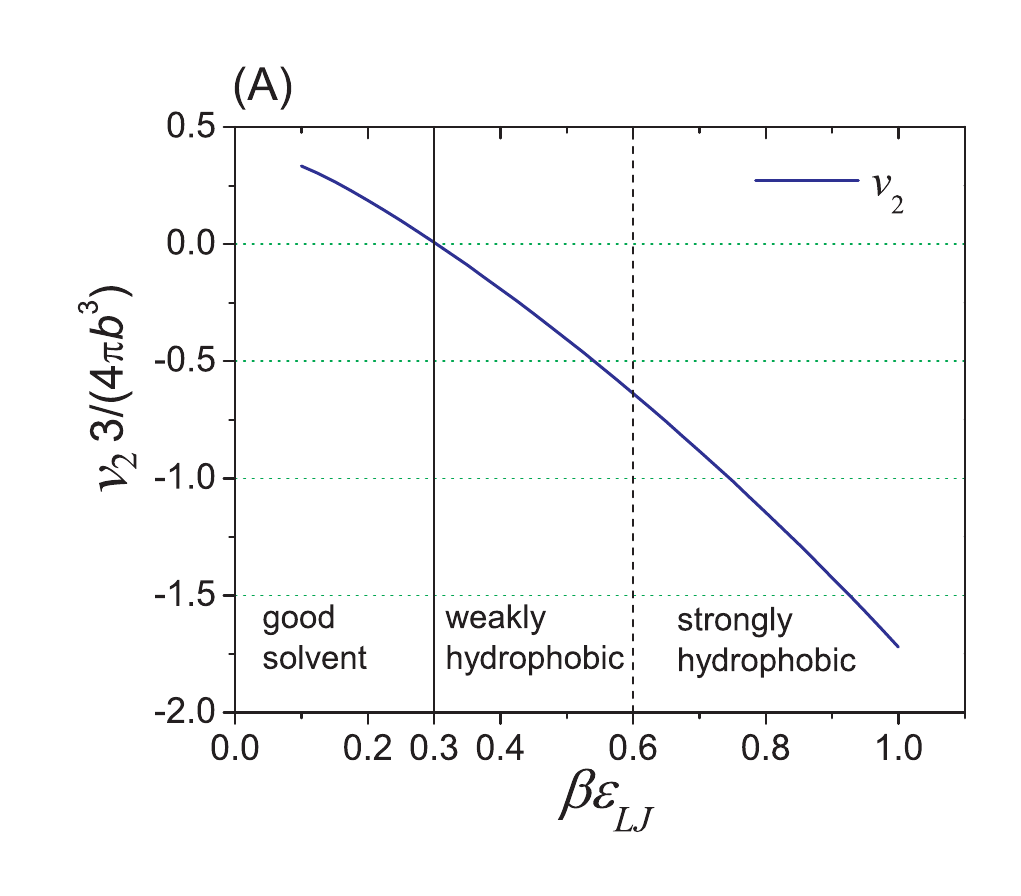}\\
   \includegraphics[width=.8\textwidth]{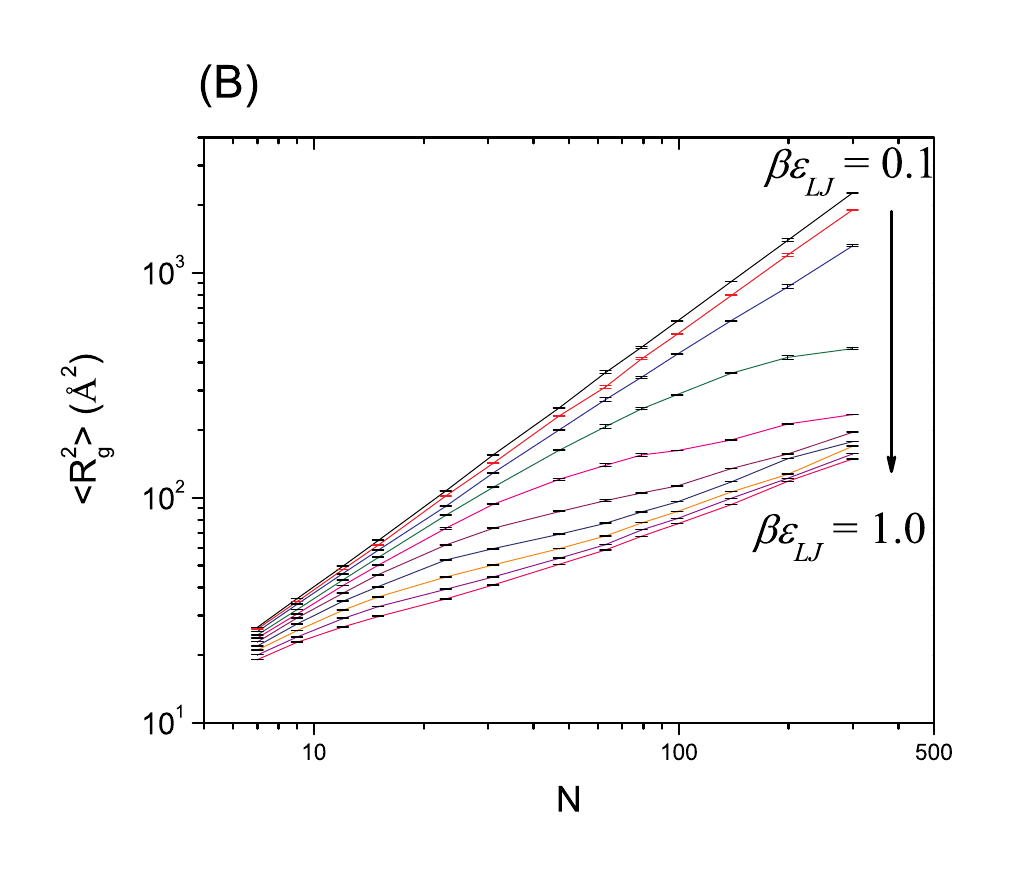}
   \caption{}
   \label{fig:virial}
\end{figure}

{\underline{Fig.  \ref{fig:virial}}}:  (A):  Second virial coefficient as a function of $\e_{LJ}$, from eq. (\ref{virial_def}).  The classification of solvent quality based on the values of $v_2$ are shown.  (B):  The variation of $\la\Rv^2_g\ra$ with $N$ for different values of $\e_{LJ}$.  The value of $\beta\e_{LJ}$ increases from 0.1 to 1.0  (in the direction of the arrow).
\\

\begin{figure}[htbp] 
   \centering
   \includegraphics[width=.48\textwidth]{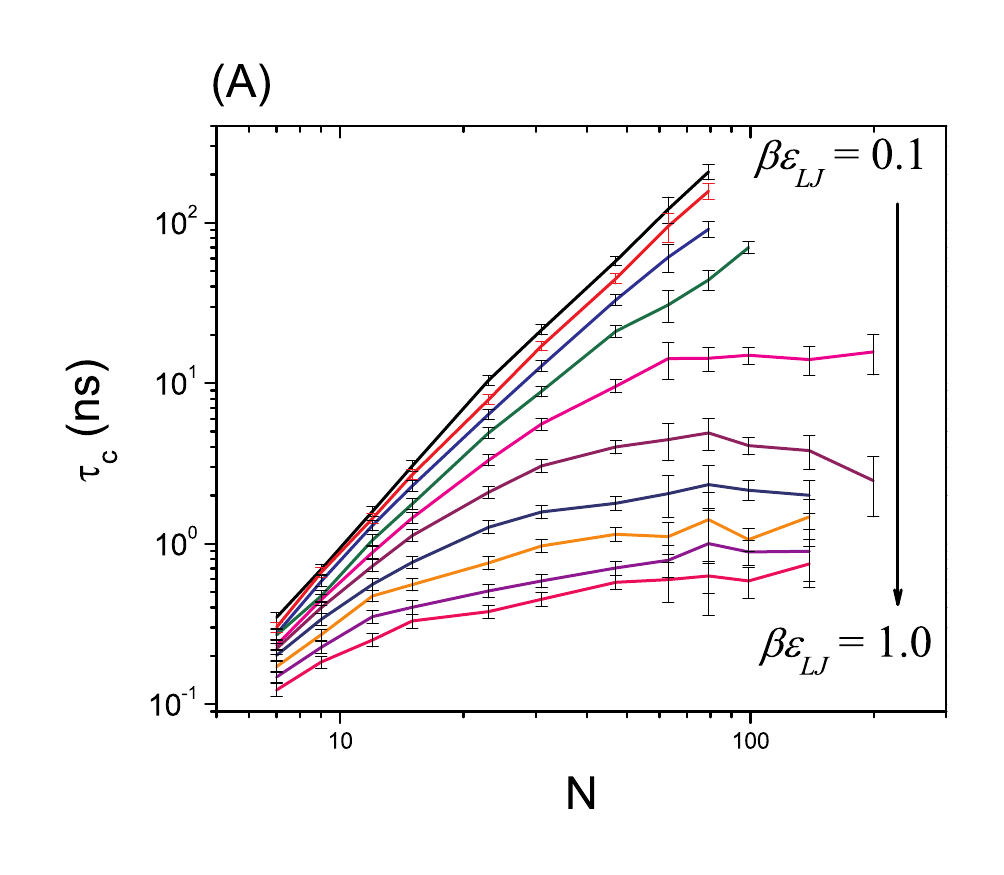}\\
      \includegraphics[width=.48\textwidth]{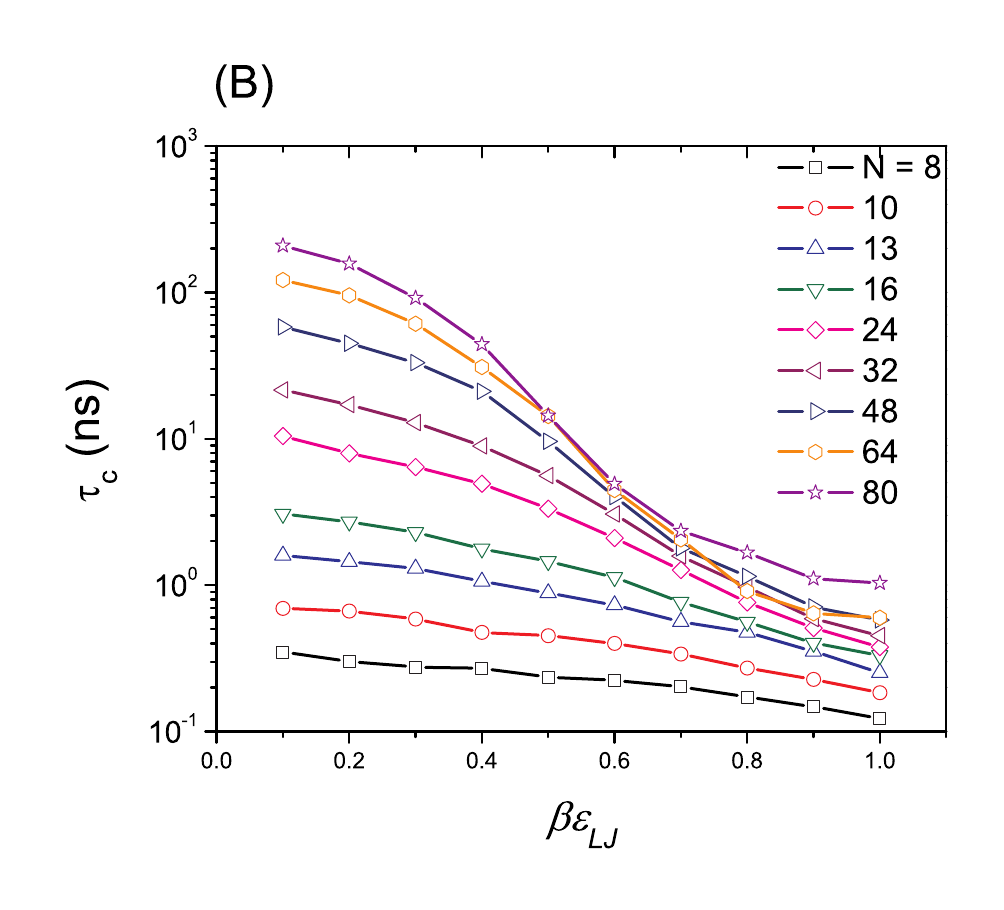}\\

   \includegraphics[width=.48\textwidth]{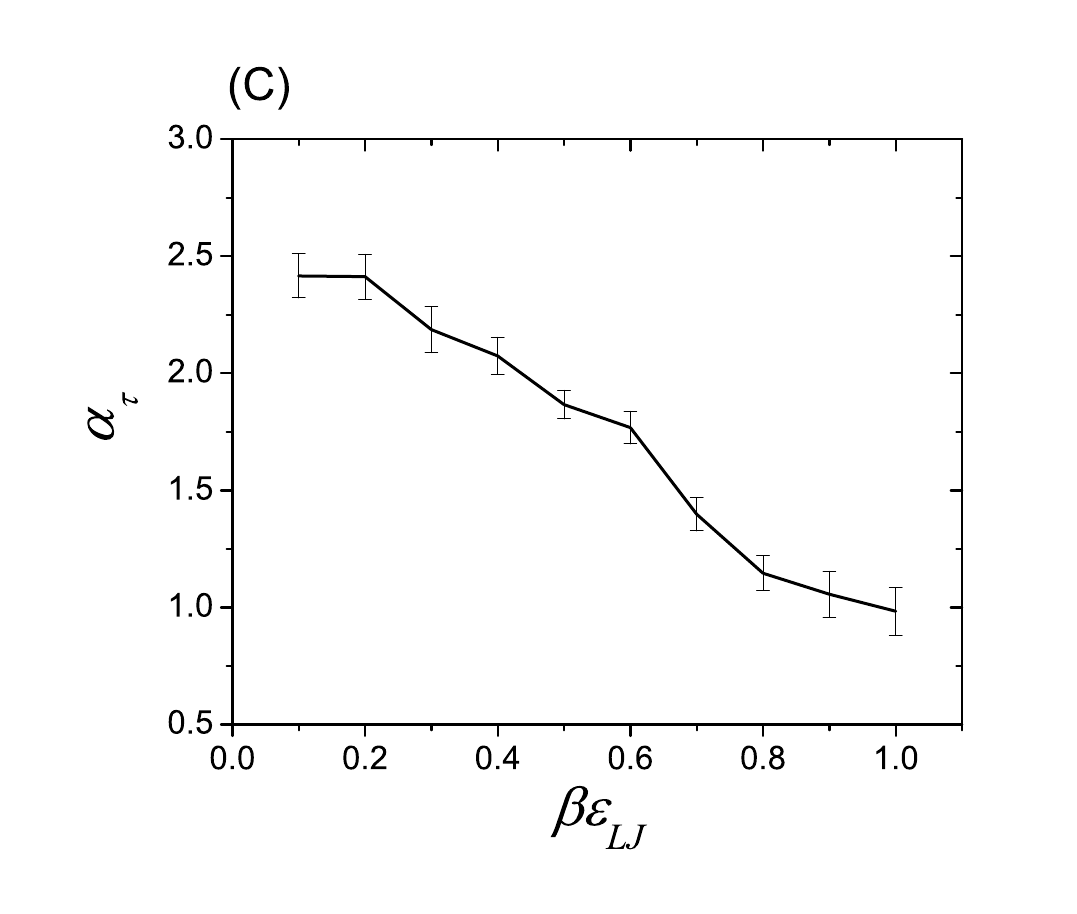}
   \caption{}
   \label{fig:tau_poor}
\end{figure}

{\underline{Fig. \ref{fig:tau_poor}}}  (A):  Loop closure time as a function of $N$ for varying solvent quality.  The values of $\beta\e_{LJ}$ increase from 0.1 to 1.0 from top to bottom, as in Fig. 3(A).  (B):  $\t_c$ as a function of $\e_{LJ}$, which is a measure of the solvent quality.  The values of $N$ are shown in various symbols.  (C):  Variation of the scaling exponent of $\t_c\sim N^{\a_\t}$ as a function of $\e_{LJ}$.
\\



\begin{figure}[htbp] 
   \centering

   \includegraphics[width=0.65\textwidth]{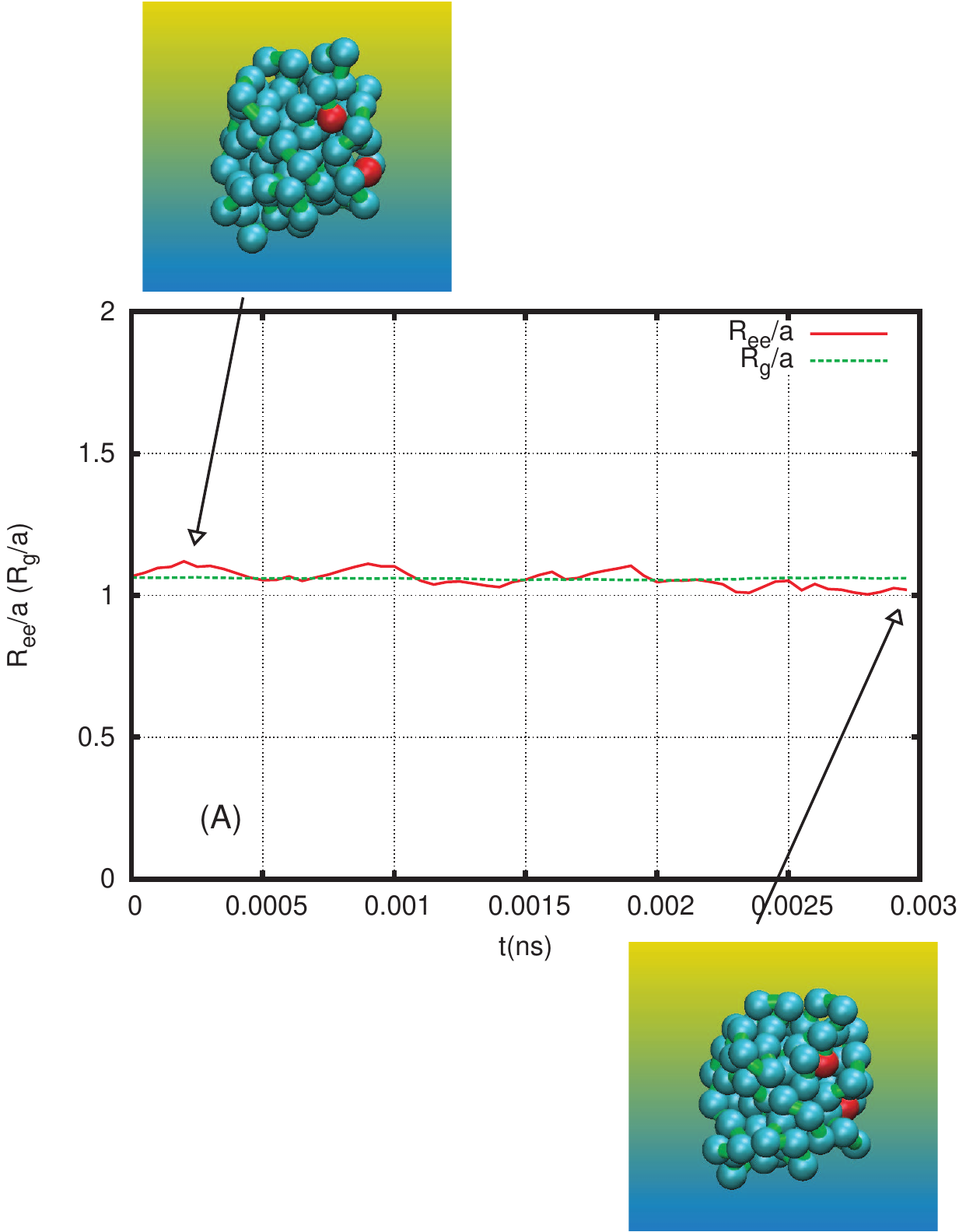}\\
\includegraphics[width=0.65\textwidth]{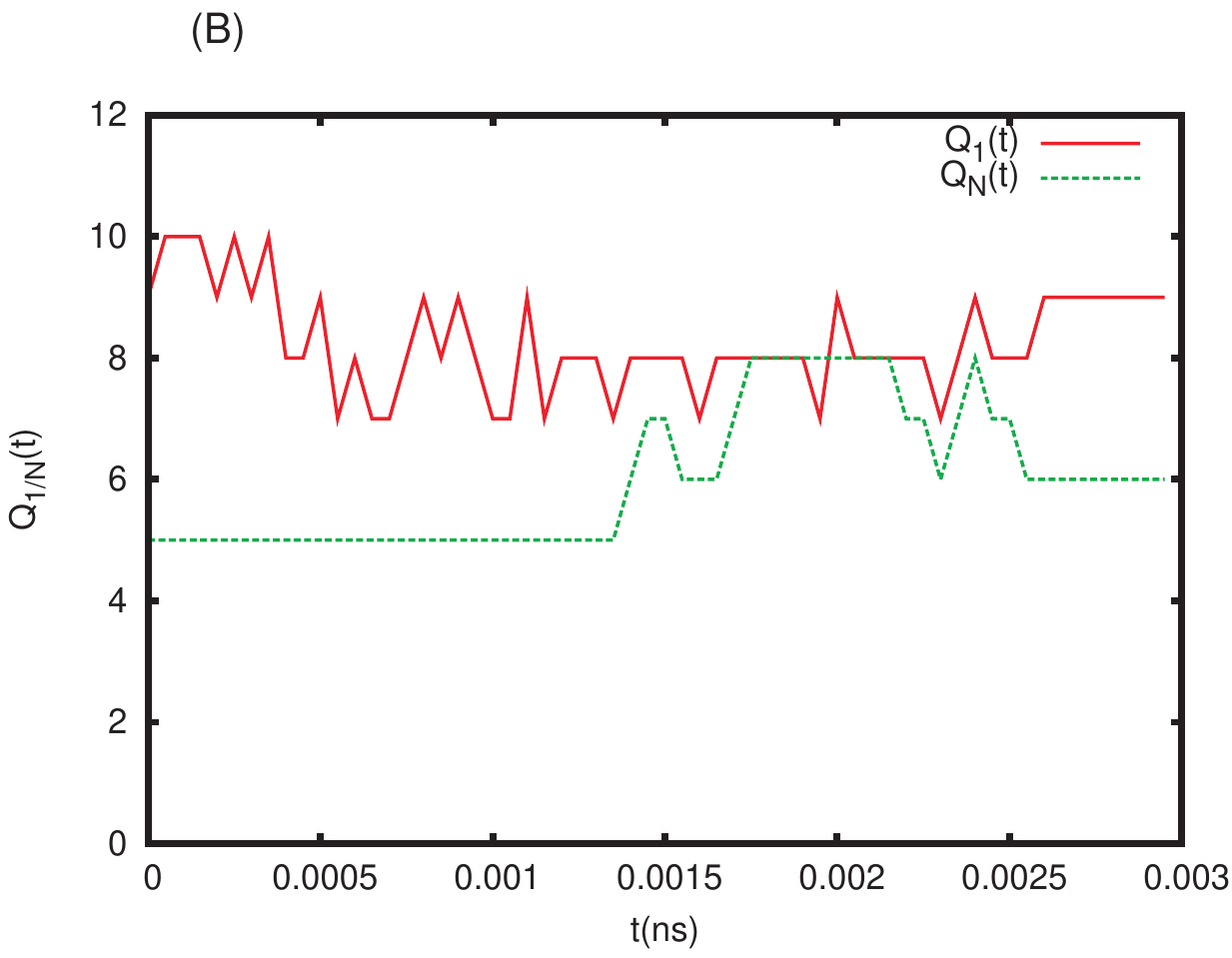}

   \caption{}
      \label{fig:Q_plot}

\end{figure}

{\underline{Fig. \ref{fig:Q_plot}}}:  Mechanism of loop closure for a trajectory with a short ($\sim 0.003$ns) first passage time.  The values of $N$ and $\beta\e_{LJ}$ are 100 and 1.0 respectively.  (A):  Plots of $|\Rv_{ee}|$ and $|\Rv_g|$ (scaled by the capture radius ($a$) as a function of time.  The structures of the globules near the initial stage and upon contact formation between the ends are shown.  The end to end distance is in red.  (B):  The time-dependent changes in the coordination numbers for the first ($Q_1(t)$) and last ($Q_N(t)$) monomers during the contact formation.
\\

\begin{figure}[htbp] 
   \centering

   \includegraphics[width=0.65\textwidth]{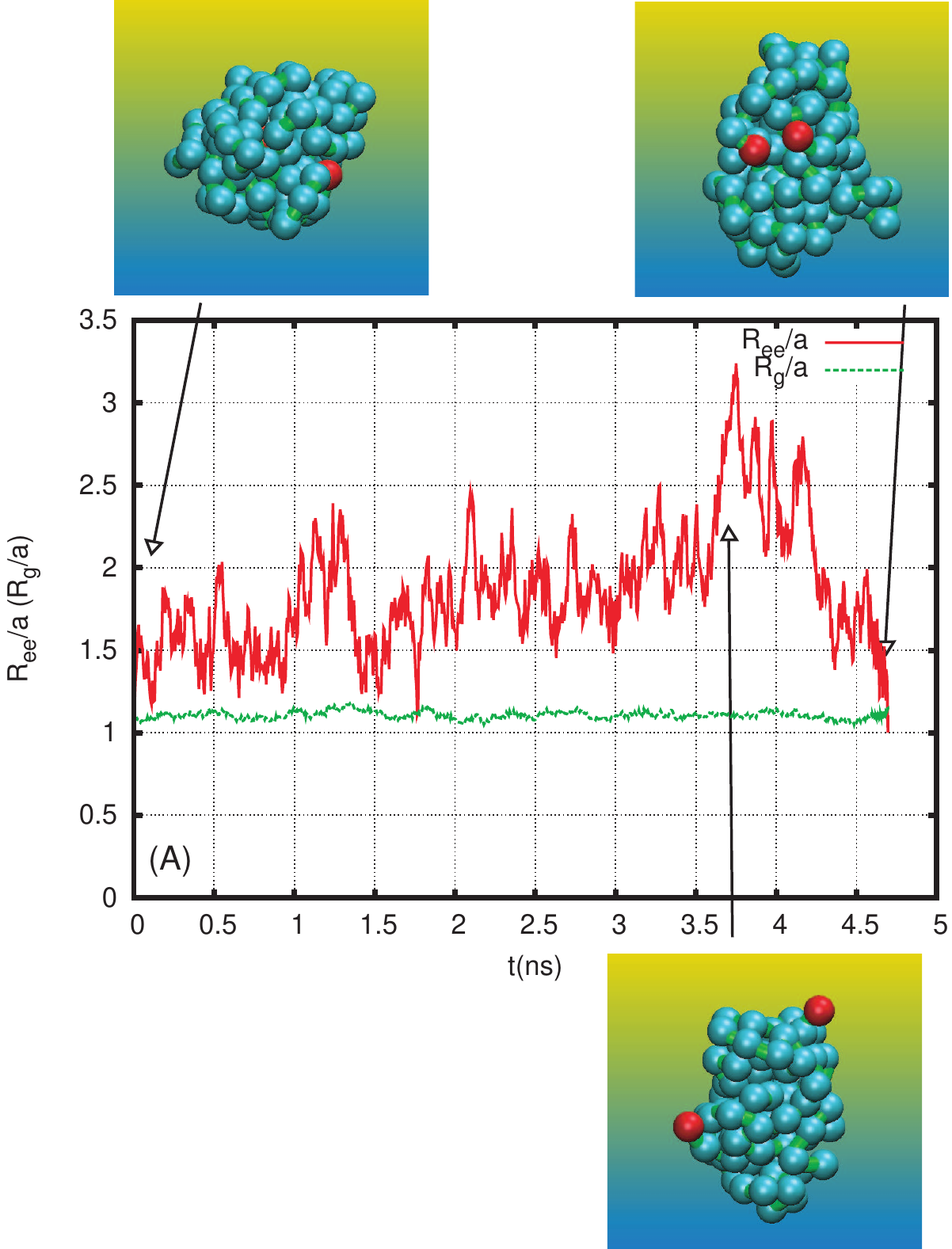}\\
\includegraphics[width=0.65\textwidth]{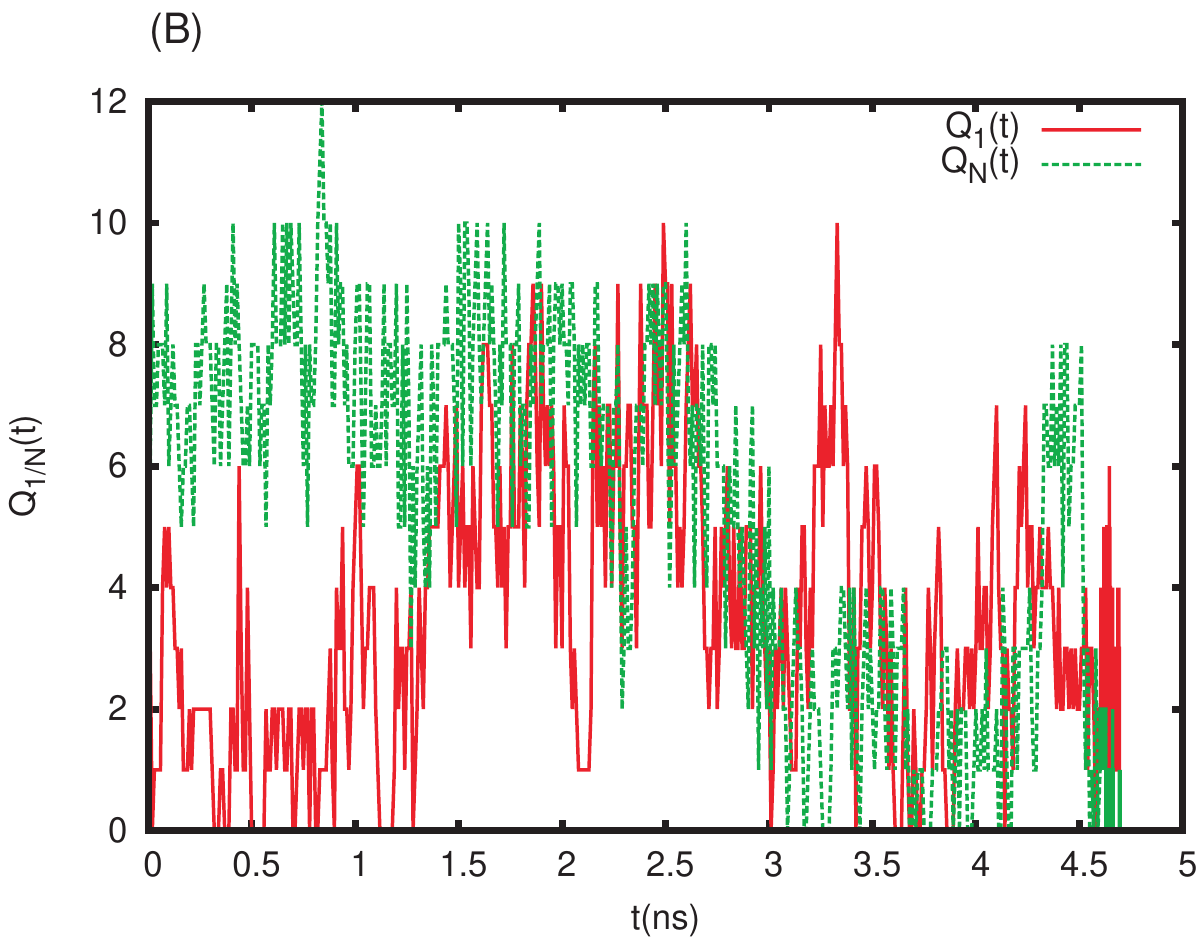}

   \caption{}
      \label{fig:Q_plot2}

\end{figure}

{\underline{Fig. \ref{fig:Q_plot2}}}:  Same as Fig. \ref{fig:Q_plot}, except the data are for a trajectory with a first passage time for contact formation that is about 4.7ns.  (A):  Although the values of $|\Rv_g|$ are approximately constant, $|\Rv_{ee}|$ fluctuations greatly.  (B):  Substantial variations in $Q_1(t)$ and $Q_N(t)$ are observed during the looping dynamics, in which both ends spend a great deal of time on the surface of the globule.
\\



\begin{figure}[htbp] 
   \centering
\includegraphics[width=0.8\textwidth]{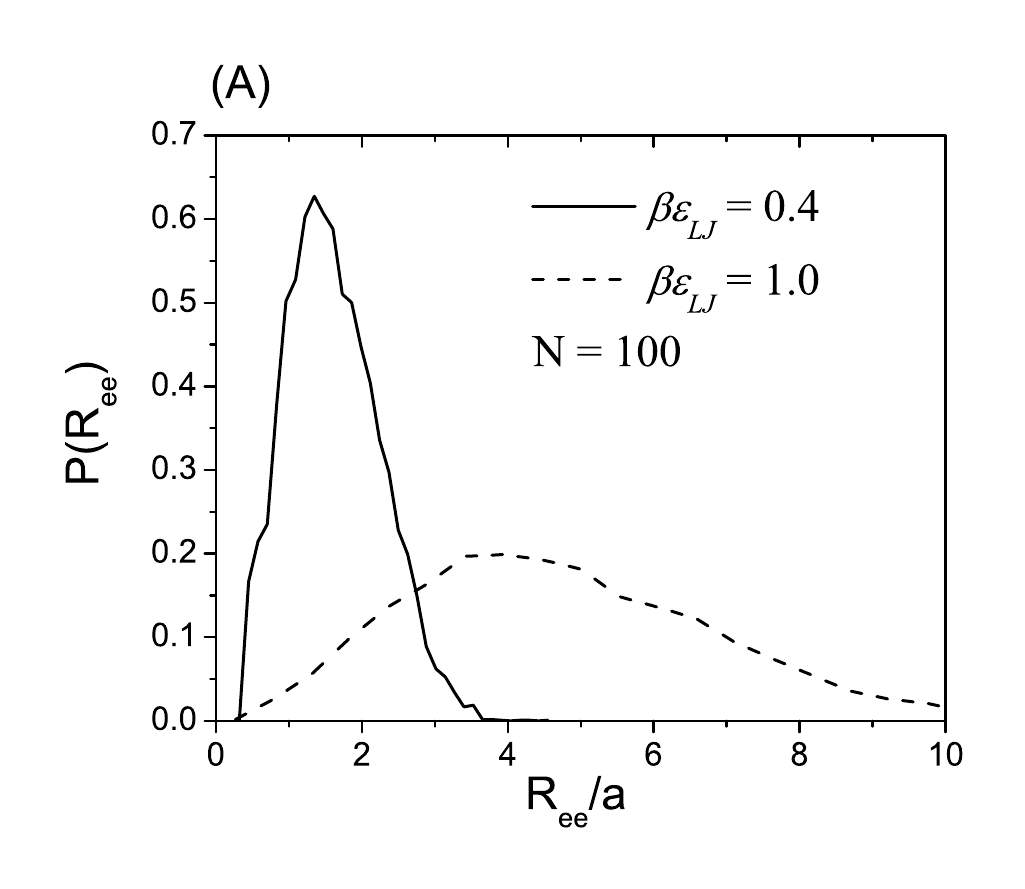}\\
  \includegraphics[width=0.8\textwidth]{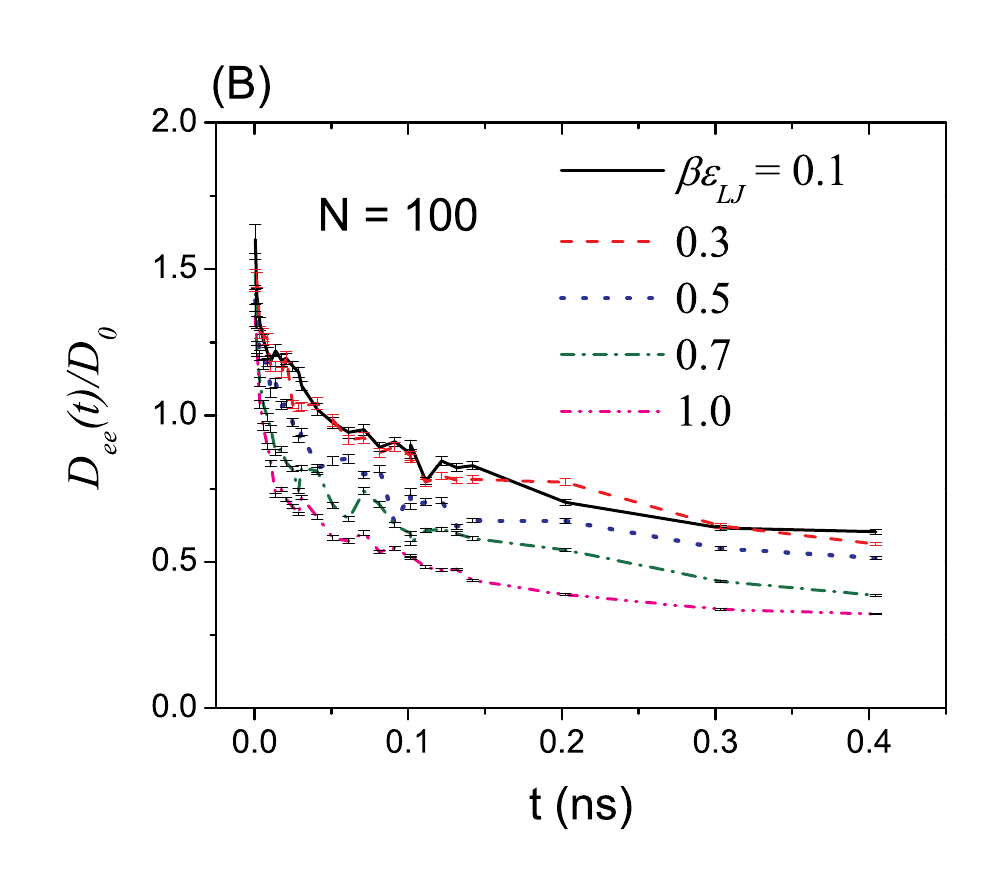}

   \caption{}
   \label{fig:pdf}
\end{figure}

{\underline{Fig. \ref{fig:pdf}}}:  (A):  Distribution of end-to-end distances for a weakly ($\beta\e_{LJ}=0.4$) and strongly ($\beta\e_{LJ}=1.0$) hydrophobic chain.  (B):  Diffusion constant $D_{ee}(t)$ in units of $D_0$ for varying solvent quality.  The diffusion constant is defined using $D_{ee}(t)=\la\d\Rv_{ee}^2(t)\ra/6t$.  The values of $\e_{LJ}$ are shown in the inset.
\\

\begin{figure}[htbp] 
   \centering
\includegraphics[width=0.8\textwidth]{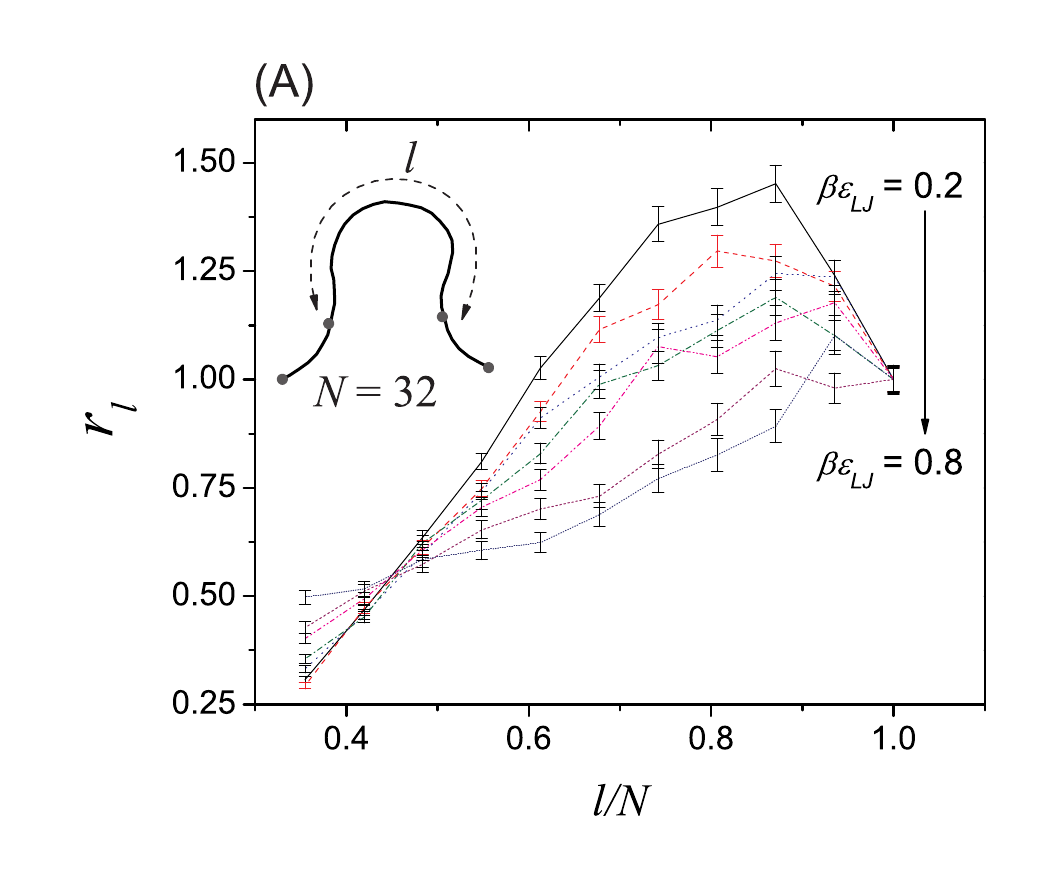}\\
  \includegraphics[width=0.8\textwidth]{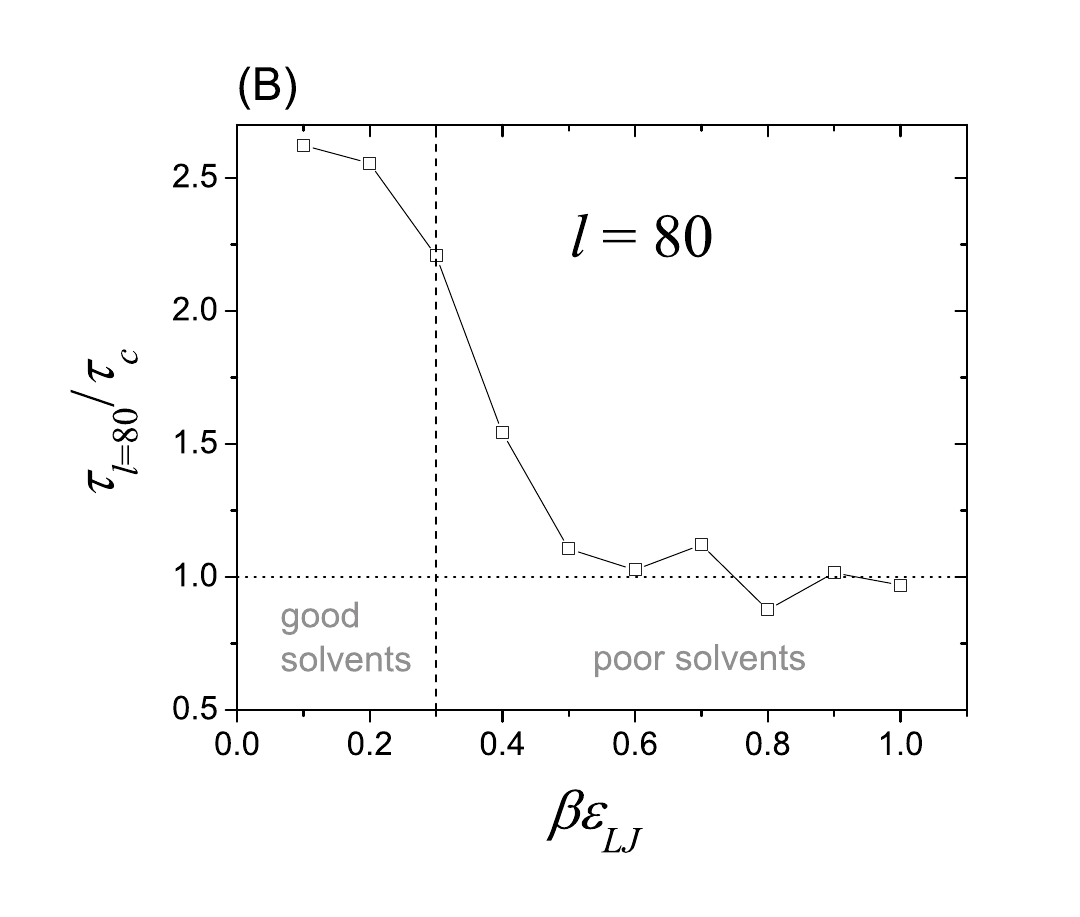}

   \caption{}
   \label{fig:rl}
\end{figure}

\begin{figure}[htbp] 
   \centering
   \includegraphics[width=.8\textwidth]{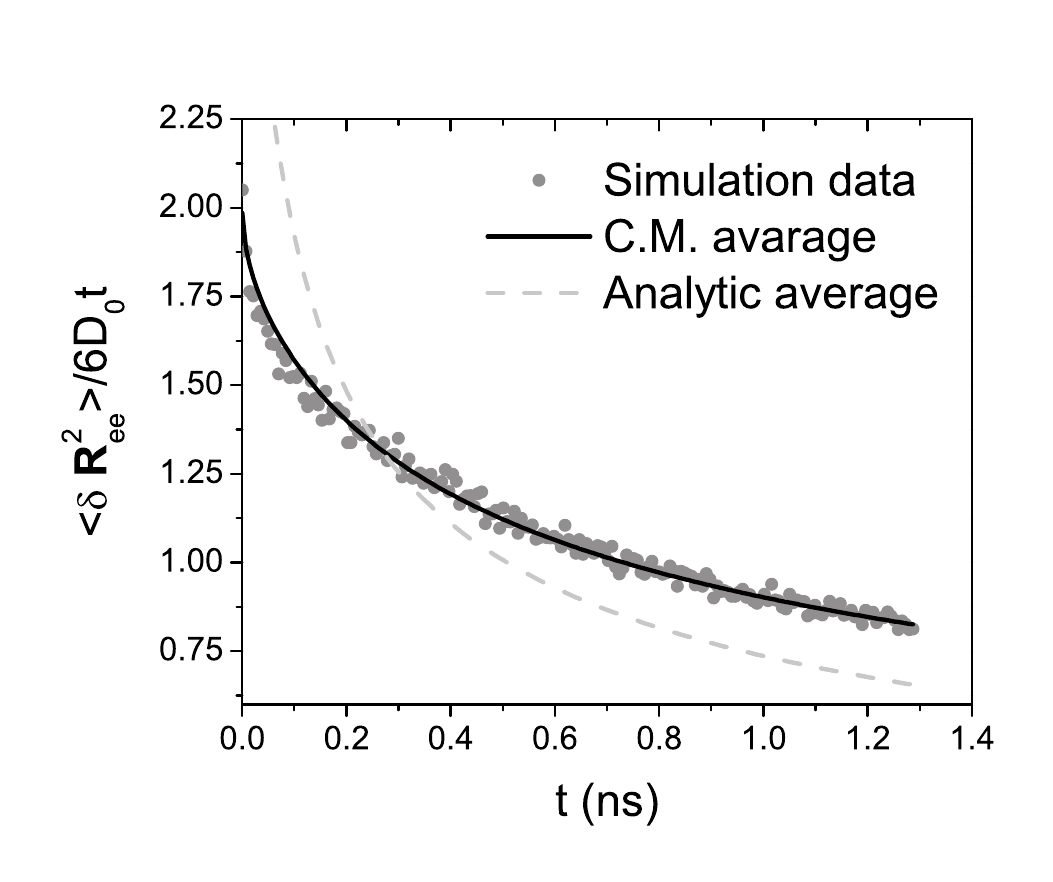}
   \caption{}
   \label{fig:diff}
\end{figure}

{\underline{Fig. \ref{fig:rl}}}: (A): The ratio $r_l=\tau_l/\tau_c$ as a function of interior
length $l$. Here $\tau_l$ is the contact formation time for beads that are separated by $l$ monomers.  $r_l$ is non-monotonic for weakly hydrophobic chains, but decreases monotonically with decreasing $l$ in the poorest solvents.  The observed maxima occur near $l/N = 0.9$. (B): For loop length $l=80$, the ratio $\tau_{l=80}/\tau_c$ as a function of $\beta \epsilon_{LJ}$ for a chain with two linkers (each of 20 beads) that are attached to beads $20$ and $100$. In good solvents, the interior loop closure kinetics is about 2.5 times slower than the end-to-end one with the same loop length. In poor solvents, however, there is virtually no difference between the two.
\\

{\underline{Fig. \ref{fig:diff}}}:  Measured Diffusion Coefficient as a function of time for the Rouse chain with $N=19$ and $b=0.39$nm.  Symbols are the simulation data, the dashed line (standard average) is obtained using eqs. (\ref{standard}) and (\ref{dt}) (with best fit $b\approx 0.26$nm), and the solid line is the center-of-mass average derived using eqs (\ref{Ree}) and (\ref{dt}) (with best fit $b\approx 0.41$nm).

\eject

%


\end{document}